\pdfoutput=1

\documentclass[prl,twocolumn,superscriptaddress,floatfix,preprintnumbers]{revtex4-1}
\pdfoutput=1
\usepackage{amsmath}
\usepackage{amssymb}
\usepackage{graphicx}
\usepackage{epsfig}
\usepackage{braket}
\usepackage{slashed}
\usepackage{bm}
\usepackage[usenames,dvipsnames]{color}
\usepackage{color}
\usepackage{mathtools}
\usepackage{pgfplots}
\usepackage{verbatim}   
\usepackage{color}      
\usepackage{subfigure}  
\usepackage{hyperref}   
\begin{document}
\title{Condensation-Driven Phase Transitions in Perturbed String Nets}
\author{Micha\"{e}l Mari\"{e}n}
\affiliation{Department of Physics and Astronomy, Ghent University, Krijgslaan 281, S9, 9000 Gent, Belgium}
\author{Jutho Haegeman}
\affiliation{Department of Physics and Astronomy, Ghent University, Krijgslaan 281, S9, 9000 Gent, Belgium}
\author{Paul Fendley}
\affiliation{All Souls College and Rudolf Peierls Centre for Theoretical Physics,  University of Oxford, 1 Keble Road, Oxford, OX1 3NP, United Kingdom}
\author{Frank Verstraete}
\affiliation{Department of Physics and Astronomy, Ghent University, Krijgslaan 281, S9, 9000 Gent, Belgium}
\affiliation{Vienna Center for Quantum Science and Technology, Faculty of Physics, University of Vienna, Boltzmanngasse 5, 1090 Vienna, Austria}

\begin{abstract}
We develop methods to probe the excitation spectrum of topological phases of matter in two spatial dimensions. Applying these to the Fibonacci string nets perturbed away from exact solvability, we analyze a topological phase transition driven by the condensation of non-Abelian anyons. Our numerical results illustrate how such phase transitions involve the spontaneous breaking of a topological symmetry, generalizing the traditional Landau paradigm. The main technical tool is the characterization of the ground states using tensor networks and the topological properties using matrix-product-operator symmetries. The topological phase transition manifests itself by symmetry breaking in the entanglement degrees of freedom of the quantum transfer matrix. 
\end{abstract}

\maketitle

\noindent\emph{Introduction}---%
The understanding and precise modelling of the entanglement structure 
is a central tool in analyzing strongly correlated quantum many-body systems \cite{hastings2007area,verstraete2006matrix,verstraete2008matrix}.
The specific entanglement pattern in ground states of local, gapped Hamiltonians for such systems led to the introduction of tensor networks, such as matrix product states (MPS) \cite{fannes1992finitely,schollwock2011density}, projected entangled pair states (PEPS) \cite{verstraete2004renormalization} and the multiscale entanglement renormalization ansatz (MERA) \cite{vidal2008class}. Substantial progress in both theoretical and numerical directions has ensued.

In this Letter, we develop such methods to give insight into an area of intense current study, non-Abelian topological phases.  Many or perhaps all two-dimensional bosonic non-chiral topological phases are characterized by the string-net models \cite{levin2005string}, whose Hamiltonians are constructed to be exactly diagonalizable. It has been proved that that the topological order survives perturbing string nets into models with non-trivial dispersion relations \cite{Klich2010,Bravyi:2010}. However, since the interesting physics requires strong correlations, its analysis away from exactly solvable points is difficult for obvious reasons.

We develop new methods to study such models, focusing 
on the doubled Fibonacci phase, the simplest anyon theory that is universal for quantum computation \cite{wang2010topological}. We provide strong evidence that the simpler Hamiltonian proposed in \cite{fendley2008topological} does indeed belong to this phase, going beyond the exact diagonalization results of \cite{PhysRevLett.110.260408} to find the anyonic excitations and their dispersion relations. Our analysis also confirms the existence of a novel quantum critical point separating this non-Abelian topological phase from a trivial one \cite{Fendley:2005,fendley2008topological}.  Combining these results therefore corroborates the general theory of anyon-condensation-driven phase transitions via spontaneous breaking of a quantum-group symmetry \cite{bais2002broken,bais2003hopf,bais2009condensate}.

Our method extends that of Ref.\ \onlinecite{haegeman2015shadows} for deformations of the toric code to the much trickier non-Abelian case. From the parametrization of the ground state as a PEPS one can construct a one-dimensional ``quantum transfer matrix'' \cite{cirac2011entanglement,zauner2015transfer,bal2015matrix}, whose fixed-point subspace 
contains the relevant features of the entanglement structure of the ground state \cite{schuch2013topological,haegeman2015shadows,liu2015environment}. This transfer matrix is a manifestation of the holographic bulk-boundary correspondence \cite{li2008entanglement,cirac2011entanglement}, and its utility does not stop at the ground state: 
from the other eigenvectors of the transfer matrix one can extract information about the elementary excitations as well 
\cite{zauner2015transfer}. As there are few techniques available to deal variationally with excitations in more than one dimension (but see \cite{vanderstraeten2015excitations} for recent progress), this insight is valuable for the understanding of the dispersion relation of two-dimensional systems. 

\noindent\emph{PEPS methodology}--- The PEPS description of general string-net wave functions has been established in \cite{buerschaper2009explicit,gu2009tensor}. We use the more recent representation introduced in the framework of matrix product operator (MPO) injectivity \cite{schuch2010peps,sahinoglu2014characterizing}, as it is especially suited to study and classify the excitations in the different possible anyon sectors. In particular, it was shown in Ref.~\cite{bultinck2015anyons} that idempotents can be constructed on the virtual level of the PEPS network that determine the different anyon sectors. We give an overview of the methodology here, and present a more detailed account in the Supplementary Material.

The Fibonacci phase has only one non-trivial anyon labelled $\tau$ in addition to the trivial identity particle, so the chirally symmetric doubled phase has four types,
labelled $1,\tau,\overline{\tau}, \tau\overline{\tau}$.  In the framework of MPO-injective PEPS, the doubled phase is characterized by two MPOs labeled by $O_1,O_{\tau}$  satisfying the Fibonacci fusion rules $O_{\tau}O_{\tau} = O_1+O_{\tau}$ \cite{sahinoglu2014characterizing}. The projector $P$ onto the vacuum sector defines the local PEPS tensor of the ground state at the renormalization group (RG) fixed point, the string-net model. It is
$$P = \frac{1}{1+\phi^2}O_1+\frac{\phi}{1+\phi^2}O_{\tau},$$
with $\phi$ the golden ratio. Acting on this tensor with a non-unitary gate on the physical level gives a perturbed PEPS no longer at the RG fixed point. Varying the gate then can drive the state through a phase transition. Below, we provide a detailed investigation of various perturbations.

\begin{figure}[h!]
\centering
$$
\vcenter{\hbox{
   \includegraphics[height=0.14\textheight]{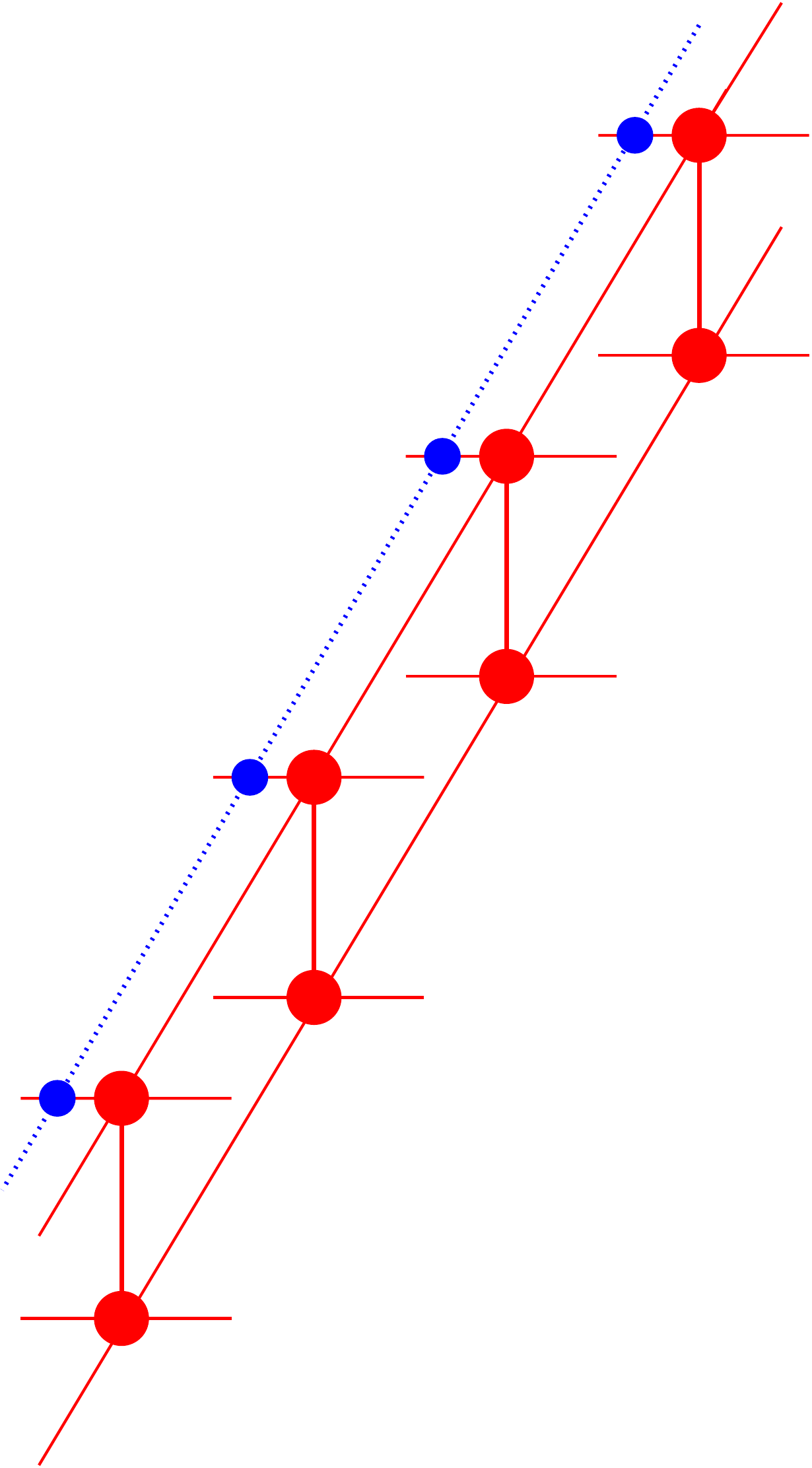}
   }}=
   \vcenter{\hbox{
    \includegraphics[height=0.14\textheight]{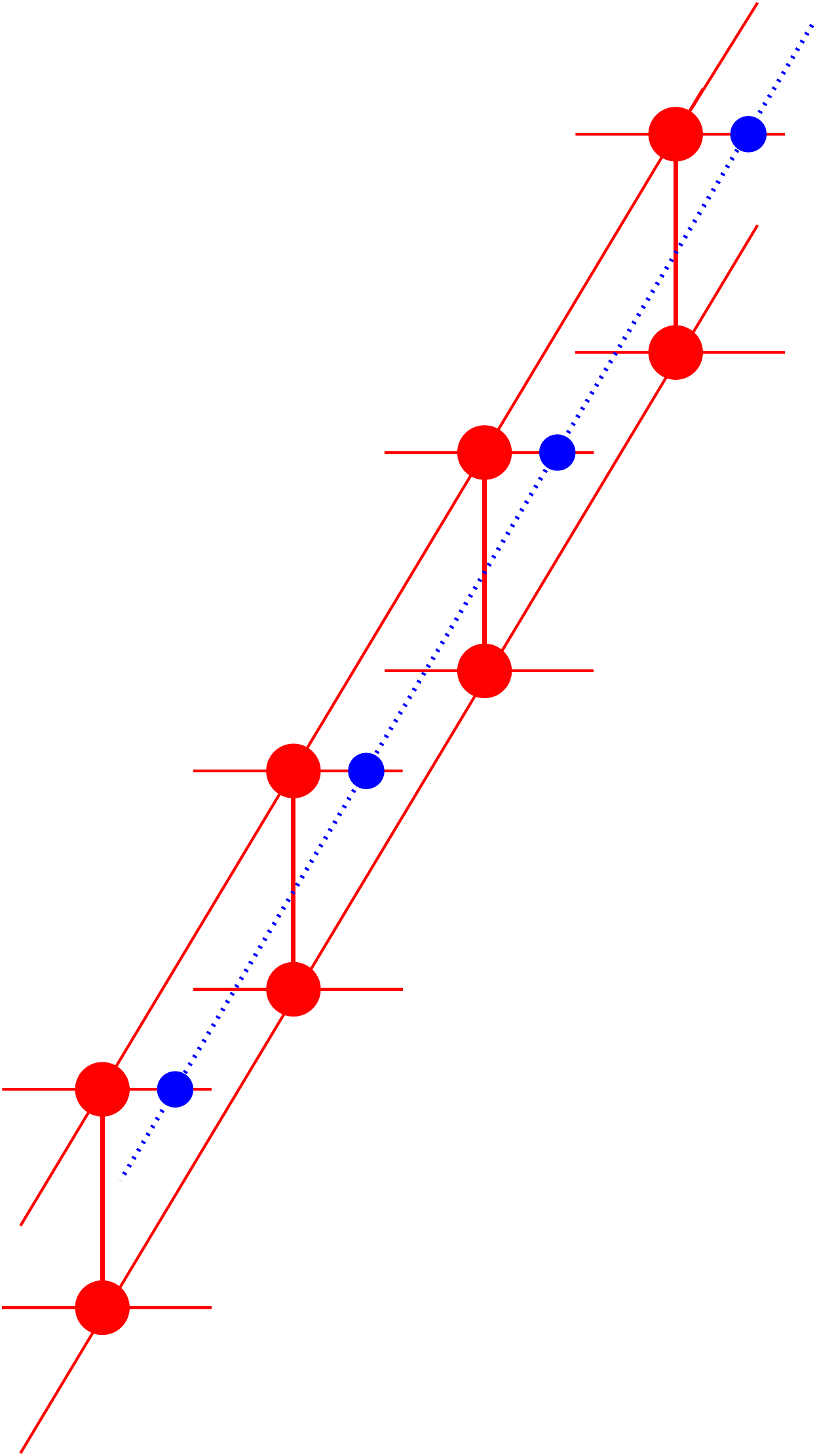}
   }}
$$
\caption{The fundamental symmetries of the transfer matrix (red) are given by MPOs (blue) that commute both on the ket and bra layer.}
\label{fig:TransferMatrixMPO}
\end{figure}

From the local PEPS tensor of the ground state, we can construct the associated quantum transfer matrix. It is a completely positive map that acts as
$$\rho \to  \sum_{\mathbf{i}} A^{\mathbf{i}} \rho (A^{\mathbf{i}})^\dagger$$
with $A^{\mathbf{i}}$ one row or column from the PEPS network and $\mathbf{i}$ the collection of physical indices on that row or column. For fixed $\mathbf{i}$, $A^{\mathbf{i}}$ is a MPO for which the local tensors have both physical and virtual dimension equal to $D$, the bond dimension of the PEPS. Likewise, the transfer matrix itself can be encoded as a MPO with dimensions $D^2$, because of the double layer (ket and bra) structure. Most important to notice is that the the individual operators $A^{\mathbf{i}}$ commute with the MPOs $O_1$ and $O_{\tau}$, so that the transfer matrix commutes with the application of these MPOs on the ket or bra layer separately as shown in Figure \ref{fig:TransferMatrixMPO}. These MPOs therefore can be thought of as creating topological defects, in the same fashion as those in two-dimensional classical lattice models \cite{Aasen:2016}. We refer to this property as the symmetry of the transfer matrix, although the MPOs $O_1$ and $O_{\tau}$ are not unitary operators. In other languages, this is called topological symmetry \cite{Feiguin:2006}, or quantum-group symmetry \cite{bais2009condensate}.

The manner in which the fixed-point subspace of the PEPS transfer matrix respects or breaks the symmetry encodes information about the condensation or confinement of the physical anyon excitations, as explained in Ref.~\cite{haegeman2015shadows} for abelian anyons. When passing through a phase transition, the fixed-point structure will change.
At the RG fixed point, we can analytically construct two fixed points and write down explicit tensor network representation for them. One is obtained from the MPO $O_1$, and the other from $O_{\tau}$. It turns out that these are the only two fixed points, as is expected. Indeed, this is the generic signature of the fact that none of the anyons has condensed or is confined. After perturbing the state, the two-fold degenerate fixed point subspace persists as long as we are in the topological doubled Fibonacci phase. There is a natural choice for two specific fixed points $\rho_1$ and $\rho_\tau$ in the subspace, which correspond exactly to `symmetry broken' states under the symmetry of the transfer matrix. 
The difference with the Abelian case is that under the action of an MPO, a fixed point can be mapped to a sum of several other fixed points, consistent with the Fibonacci fusion rules $\tau \times\tau = 1  + \tau$. The full rules for the result of applying MPOs to the fixed point are as expected from the notation,
\begin{align*}
O_{\tau}\rho_1&= 
\rho_1 O_{\tau}^{\dagger}=\rho_{\tau}, \quad 
&&O_{\tau}\rho_1 O_{\tau}^{\dagger}=\rho_1+\rho_{\tau}
\\
O_{\tau}\rho_{\tau}&=\rho_{\tau}O_{\tau}^{\dagger}=\rho_1+\rho_{\tau},\quad
&&O_{\tau}\rho_\tau O_{\tau}^{\dagger}=\rho_1+2\rho_{\tau}.
\end{align*}
The two fixed points $\rho_1$ and $\rho_\tau$ furthermore have the property that they are injective as MPS, and are therefore exactly the ones that are approximated by numerical MPS algorithms. 

Having obtained both fixed points, we can construct an approximation to the `excitations' of the transfer matrix, i.e.\ the eigenvectors corresponding to the next eigenvalues $\mu=\mathrm{e}^{-\lambda}$ of largest magnitude (smallest real part of $\lambda$), using the ansatz discussed in \cite{haegeman2015shadows,PhysRevB.85.035130,PhysRevB.85.100408}. 
One option is to construct excitations by locally changing a tensor of  $\rho_1$ or $\rho_{\tau}$. The second option is to construct domain-wall excitations by using both fixed points, one on each side of the locally changed tensor. We can easily make momentum eigenstates by taking the superposition of translations of such excitations with appropriate phases and hence obtain a `dispersion relation' of the transfer matrix.  In order to classify these excitations in terms of the physical anyon sectors, we need to reinterpret them in the context of the ``mixed'' transfer matrix. In particular, we rewrite the domain-wall excitations to replace the kink with a half-infinite extra MPO attached to the site with the perturbed tensor. Due to the non-Abelian character of the theory, there are several possibilities to obtain this MPO string from a kink excitation. 
To create real anyon excitations we also need to form very specific combinations of the variational states in every given energy and momentum sector. These combinations correspond exactly to excitations that are constructed on top of the minimally entangled ground states \cite{zhang2012quasiparticle}. We then use the virtual idempotents from Ref.~\cite{bultinck2015anyons} to obtain a complete classification.

We expect all phase transitions to be towards the topologically trivial phase because of the general arguments of Ref.~\cite{bais2002broken,bais2003hopf,bais2009condensate}. With our approach, we find four different fixed points of the transfer matrix, which we can indeed interpret as the confinement of the $\tau$ and $\overline{\tau}$ anyons, driven by the condensation of the $\tau\overline{\tau}$ anyon. More specifically, as the fixed point $O_{\tau}\rho_1 O_{\tau}^\dagger$ is now orthogonal to the fixed point $\rho_1$,  the presence of MPO strings is suppressed and the `energy' associated with a pair of $\tau$ or $\overline{\tau}$ anyons increases strongly with their separation distance. More details are provided in the explicit results below.

\noindent\emph{Results}---We focus now on several perturbations of the Fibonacci string-net ground state and study the dispersion relation of the resulting states. As a full variational method to find ground states of perturbed string-net Hamiltonians is currently out of reach, we instead perturb at the level of the state by extending the filtering procedure introduced in \cite{Ardonne:2003,PhysRevB.77.054433} using the framework of PEPS. 
We stress that each of the perturbed states is still the ground state of a local Hamiltonian. Indeed, we always start from the Fibonacci string net PEPS $\ket{\Omega}$ with the corresponding positive Hamiltonian $H=\sum_v A_v + \sum_p B_p:=\sum_j h_j$ that consists of commuting plaquette and vertex terms \cite{levin2005string}.  All perturbed states we consider are of the form $\prod_i Q_i \ket{\Omega}$ with $Q_i$ a local and positive operator on site $i$ and clearly these states are ground states of the local Hamiltonian $\sum_j Q^{-1}(j)h_jQ^{-1}(j)$ with $Q(j)$ the product of all $Q_i$ with the site $i$ in the support of $h_j$. 

The Fibonacci string-net model is defined by putting a two-state quantum system (a qubit) on each edge $j$ of some lattice. Each segment of string net corresponds to $\sigma_j^z=1$ acting on the corresponding qubit. The original string-net ground state is a sum over configurations with no ``ends'' (i.e.\ no vertices with only one segment of string-net touching). A key fact is that the weight of each configuration in the ground state depends only on topological data \cite{Fidkowski:2009}. Equal-time correlators in the ground state are thus the same as those in a corresponding 2D classical system in the Rokhsar-Kivelson fashion \cite{Rokhsar:1988}. As described in detail in Ref.\ \onlinecite{fendley2008topological}, the string-net model on the honeycomb lattice is related to the classical 2D ferromagnetic $q$-state Potts model with $q=2+\phi$ on the dual triangular lattice at infinite temperature. 

The first perturbation we consider is a string tension. We act with a local $Q_j=\exp(-\beta \sigma^z_j)$ operation on every qubit. 
Increasing $\beta$ from 0 still gives the ground state as a sum over local nets without ends, but with a local weighting that favors shorter nets. This simply corresponds to lowering the temperature in the Potts model, so the phase transition in the quantum theory therefore can be located from the classical theory. The Potts models with $q\le 4$ have a second-order phase transition at a finite temperature, which for $q=2+\phi$ on the triangular lattice is $\frac{1}{4}\log(x \sqrt{\phi+2}+1)$ with $x$ the positive root of $\sqrt{\phi+2}x^3+3x^2+1=0$ \cite{wu1982potts,baxter1978triangular}. The corresponding 2D conformal field theory (CFT) describing the scaling limit is the minimal rational one labelled $(9,10)$ \cite{francesco2012conformal}. It is worth noting that the identical CFT describes the scaling limit of the quantum critical Fibonacci ladder \cite{gils2009topology}.   

\begin{figure*}
\centering
\includegraphics[width=0.98\textwidth]{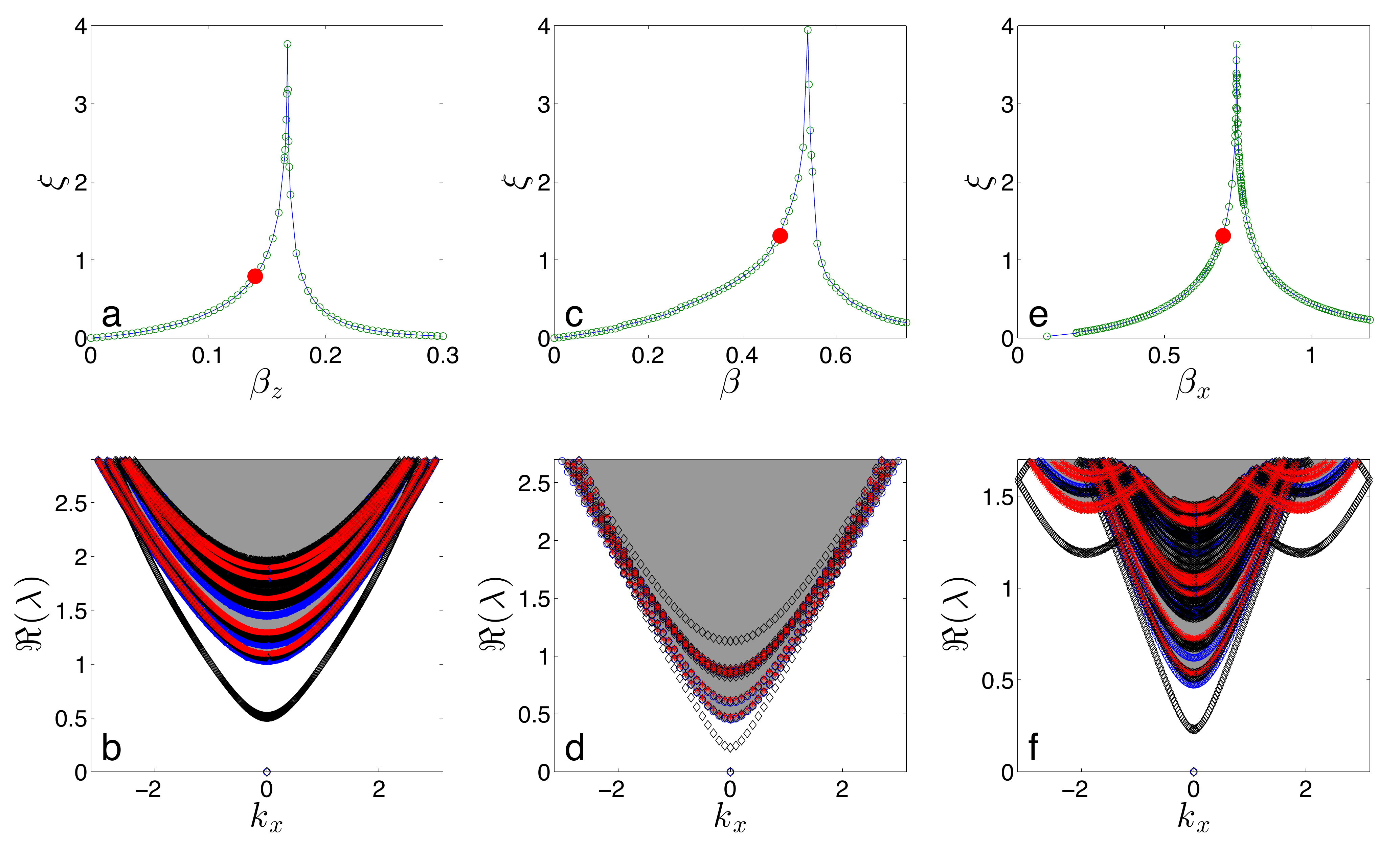}
\caption{
The correlation length of the $\exp(-\beta_zZ)$ interpolation (a), an interpolation to the Fendley model (c) and  $\exp(-\beta_x X)$ filtering (e). The points where the dispersion relations are given are shown in red. The dispersion relation of the string net model with $\exp(-\beta_zZ)$ filtering at $\beta_z=0.14$ (b), the Fendley model (d) and $\exp(-\beta_x X)$ filtering at $\beta_x=0.7$ (f).}
\label{fig:all}
\end{figure*}

The perturbed string net exhibits a quantum phase transition at the corresponding string tension, and
the quantum transfer matrix arising from PEPS shows the structure beautifully.  Indeed, this transfer matrix can be mapped to that of the $2+\phi$-state Potts model.  From it we can easily extract the correlation length of the state numerically. We confirm that as predicted, a critical phase transition to a trivial state occurs for $\beta \approx  0.16776$, as illustrated in Figure \ref{fig:all}(a). As a check on the methods, we obtained the same location of the phase transition using fidelity methods (not shown) \cite{PhysRevA.78.010301}. Beyond the critical point, by further increasing $\beta$, the state is not topologically ordered anymore, as is indeed reflected in the fixed point structure. Beyond the phase transition, we find four different fixed points, which excludes the possibility of having anyon excitations with a non-trivial MPO string $O_{\tau}$. Such strings are then suppressed and the anyons confined.

Furthermore, no traces of bound states appear in the spectrum, which are expected in a generic quantum 2D phase transition out of the Fibonacci phase from perturbation theory \cite{private}. 

An analogous fine-tuned transition and corresponding dimensional reduction also appears in the toric code model \cite{Ardonne:2003,PhysRevB.77.054433,isakov2011dynamics}, but here the phase transition is much more intricate. Phase transition out of non-Abelian phases using fine tuned Hamiltonian interpolations where studied previously in \cite{burnell2012phase,burnell2011condensation}.
It would be interesting to recover the full CFT signature from finite-size simulations and scalings and compare these to the results for the Fibonacci ladder \cite{gils2009topology}.
However, we wish to stress that the resemblance to a 1D quantum/2D classical phase transition is a mark of the specific model considered, not of the method itself. A way to consider more generic interpolations with general PEPS that can for instance change the bond dimension of the PEPS will be considered elsewhere \cite{preparation}. We expect that to capture a genuine 2D quantum phase transition the bond dimension of the PEPS has to grow and eventually diverge. The topological information, however, is still contained in the MPO symmetries of the quantum transfer matrix and the tools used here can still be used.

The phase transition can be probed in more depth by plotting the dispersion relations of the excitations of the quantum transfer matrix. 
We consider only chirally symmetric perturbations, so the $\tau$ and $\overline{\tau}$ anyons are always exactly degenerate. 
Moreover, the condensation of $\tau$ or $\overline{\tau}$ individually is prohibited by general arguments \cite{bais2002broken,bais2003hopf,bais2009condensate}, leaving $\tau\overline{\tau}$ as the only non-trivial possibility. As the fusion product of two $\tau\overline{\tau}$ anyons contains all other anyons, the two-particle continuum of these elementary excitations has support in all topological sectors. We color this continuum grey, but also continue to plot the lowest excitations found by the algorithm in every sector. The four different anyon sectors are identified using the methods of Ref.\ \onlinecite{bultinck2015anyons}, and are color coded as  $1$ (blue circle), $\tau,\overline{\tau}$ (red crosses and plusses) and $\tau\overline{\tau}$ (black diamond). 
In Figure \ref{fig:all}(b) we plot the spectrum $\Re(\lambda) = \log \lvert \mu\rvert$ at $\beta_z = 0.14$. We clearly see a condensing $\tau\overline{\tau}$ anyon and no traces of a bound state. These observations are still valid and observable closer to the critical point. For instance, the same calculations where performed for $\beta_z=0.161$ (not shown) and gave similar results.

We now consider including a weighting for trivalent vertices in the string net, i.e.\ a penalty for configurations where three strings meet at a vertex. We can then check the argument that the net model with a simpler Hamiltonian described in \cite{fendley2008topological,PhysRevLett.110.260408} is in the same topological phase as the Fibonacci string net. 
We tune this weight to follow a path from the string-net ground state to the ``quantum self-dual'' ground state \cite{fendley2008topological}. Both the correlation length in Figure \ref{fig:all}(c), and the fidelity approach (not shown) indicate a phase transition only beyond the quantum self-dual state, confirming the suggestion of \cite{PhysRevLett.110.260408} that the loop model is in the same phase as the Fibonacci string net.  We illustrate the dispersion relation in Figure \ref{fig:all}(d). Again, we see that a $\tau\overline{\tau}$ anyon condenses. Qualitatively this phase transition is similar to the first one, because the weight for trivalent vertices in the corresponding classical model is an irrelevant perturbation. 

Finally we consider a filtering with $Q_i=\exp(-\beta \sigma^x)$. This interpolation violates the closed-net condition and creates open strings in the states, which are the hallmark of anyons. We show the correlation length of the filtered state along this path in Figure \ref{fig:all}(e). The dispersion relation is shown in Figure \ref{fig:all}(f), as expected it is again the $\tau\overline{\tau}$ anyon that condenses.

\noindent\emph{Conclusion and Outlook}---
We have shown how the dispersion relation of a perturbed 2D string-net Hamiltonian is reflected in the eigenvalues of the 1D quantum transfer matrix arising from the PEPS description of the ground state. Our results clarify how the topological properties of the ground states and the excitations are related to the symmetry properties of the PEPS transfer matrix and its fixed point spectrum,  extending the findings of Ref.\ \onlinecite{haegeman2015shadows} to non-Abelian theories. This enabled us to classify the excitations into different anyonic sectors and to find their dispersion relations. We identified the anyon type of the condensing particles at a non-trivial quantum critical point, thereby confirming abstract theoretical results \cite{bais2009condensate} in a concrete quantum-many-body system.

Our work opens up several directions for future work. First, it is an interesting question how the results of our method can quantitatively match results from perturbation theory \cite{private} by using more elaborate interpolations. 
Second, a similar procedure can be carried out for models such as the Ising string net which has a non trivial condensation driven phase transition to the Toric Code phase. It would be interesting to see how this transition is reflected in the transfer matrix.
Finally, we want to stress that the characterization of anyons using the idempotents can be applied, even more straightforward, to the full 2D setting. Hence, given a good PEPS representation of the ground state, the variational excitations can be calculated \cite{vanderstraeten2015excitations} and classified accordingly.

\begin{acknowledgements}
We acknowledge inspiring discussions with Laurens Vanderstraeten, Norbert Schuch, Julien Vidal. This work was supported by the Austrian Science Fund (FWF) through grants ViCoM and FoQuS, the EC through grants QUTE and SIQS, and EPSRC through grant EP/N01930X. J.H. and M.M. acknowledge support by the FWO. 
\end{acknowledgements}

\bibliographystyle{apsrev4-1}
\bibliography{Shadows}

\newpage
\onecolumngrid

\section{Appendix}
\subsection{Topological Order and Anyon Excitations in PEPS}%

Tensor network states known as PEPS are believed to provide a good and efficient representation for the ground states of gapped, local 2D Hamiltonians. The framework of MPO-injective PEPS is more specifically designed for the study of topologically ordered models. In this Appendix, we first give a concise review of the formalism of MPO-injective PEPS. The reason PEPS can deal with topologically ordered phases is that the tensors can encode the topological information locally in their virtual degrees of freedom. 

A generic PEPS satisfies the so called injectivity condition, after sufficient blocking the tensor is invertible as a map from its virtual to physical degrees of freedom. In contrast, the virtual space on which an MPO-injective PEPS can be inverted is not the full virtual space but only a specific subspace. This subspace is characterized by a (non-injective) MPO projector $P$. We can write $P$ as a finite linear combination of several injective MPOs $O_a$ with size independent coefficients, $P=\sum_a w_a O_a$. These MPOs now form a fusion algebra $O_aO_b=\sum_c N_{ab}^cO_c$ with integer coefficients and fully characterize the topological information in the PEPS \cite{bultinck2015anyons}.

\begin{figure}[h]
$$
\vcenter{\hbox{
\includegraphics[width=0.15\linewidth]{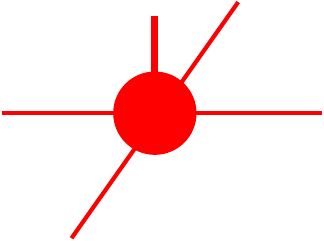}}} = 
\vcenter{\hbox{  
\includegraphics[width=0.15\linewidth]{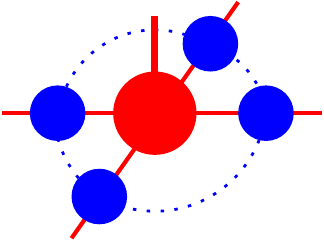}}} 
$$
\caption{The PEPS tensor $A$ is invariant under the MPO-projector (blue).          
}
\label{fig:MPOinvariant}
\end{figure}

If we start from the MPO projector $P$ or its algebra generated by the injective MPOs $O_a$, we can characterize an associated MPO-injective PEPS using the following conditions. First, the PEPS tensor is invariant under the application of the MPO  projector $P$ on the virtual space. This requirement is graphically illustrated in Figure \ref{fig:MPOinvariant}.

\begin{figure}[h!]
$$
\vcenter{\hbox{
\includegraphics[width=0.15\linewidth]{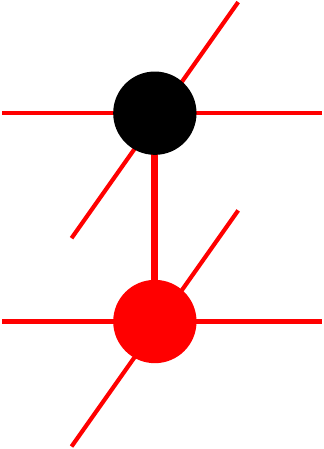}}} = 
\vcenter{\hbox{  
\includegraphics[width=0.15\linewidth]{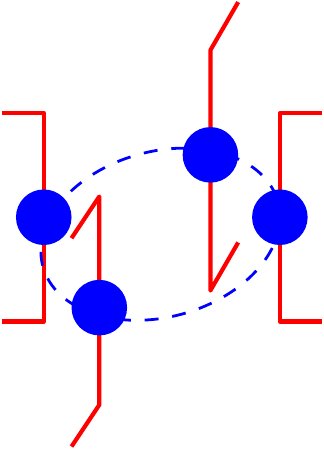}}} 
$$
\caption{MPO-injectivity requires the existence of a tensor (black) that acts as the pseudoinverse on the space determined by the MPO projector.}          
\label{fig:MPOinjective}
\end{figure}

Second, the PEPS tensor is injective as a map from the virtual to the physical space when we restrict the domain to the subspace characterized by an MPO projector, see Figure \ref{fig:MPOinjective}. 
 
\begin{figure}[h!]
\begin{align}
\vcenter{\hbox{
\includegraphics[width=0.15\linewidth]{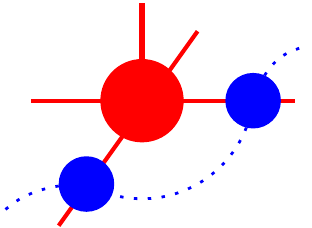}}} =
\vcenter{\hbox{
\includegraphics[width=0.15\linewidth]{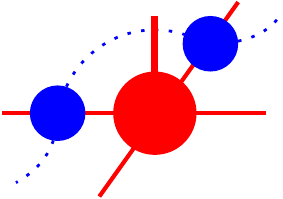}}}
\end{align} 
\caption{The pulling through property.}
\label{PullThrough}
\end{figure}

The third crucial requirement is the following. The MPO formalism is heavily based on the intuition that topological properties in 2D arise from the existence of long strings that cannot be detected locally but that can have observable nonzero winding numbers. We encode this information in local tensors by requiring that we can pull the MPOs $O_a$ through the virtual network as illustrated in Figure \ref{PullThrough}. 

\begin{figure}[h!]
\begin{align*} 
\vcenter{\hbox{
\includegraphics[width=0.35\linewidth]{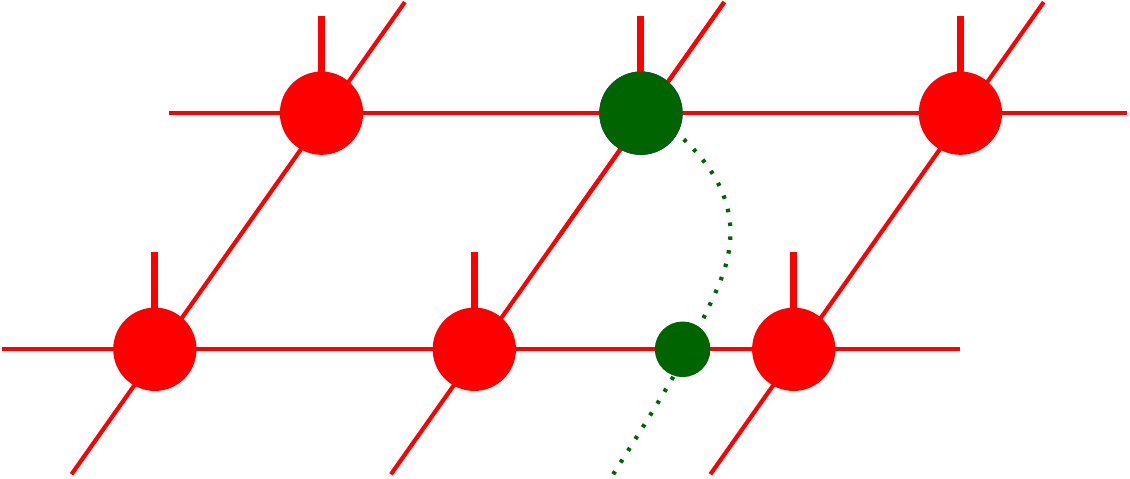}}} =
\vcenter{\hbox{
\includegraphics[width=0.35\linewidth]{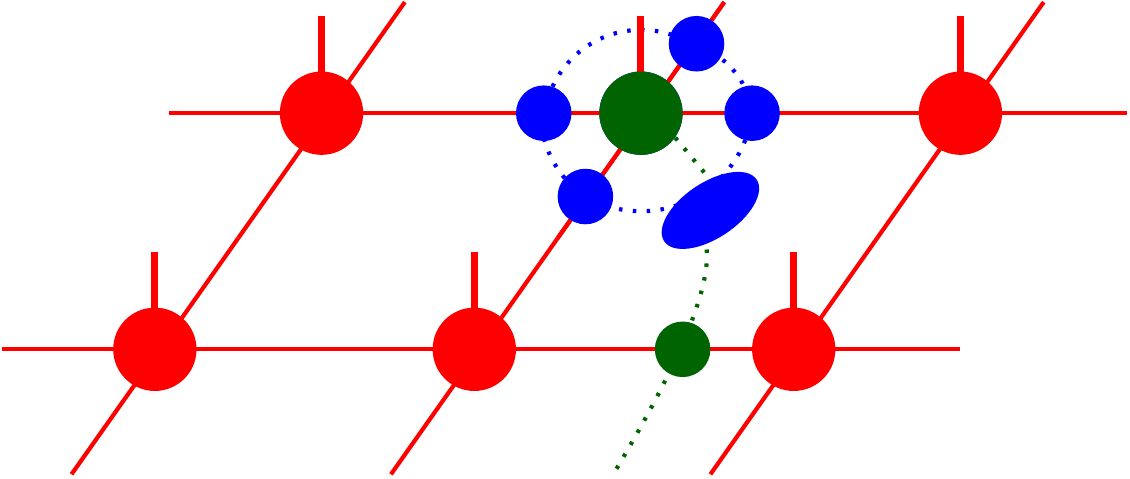}}}
\end{align*} 
\caption{An excitation, implemented in the PEPS language by the modified green tensor, is of one definite anyon type if it lives exactly in the support of one of the elementary idempotents of the theory. These idempotents are constructed by closing an MPO loop with a fixed, specific tensor (blue oval), one for each anyon type.}
\label{fig:Excit}
\end{figure}

Given a representation of the ground state of a local Hamiltonian as a PEPS, one can argue that elementary excitations can be obtained by modifying a single tensor in the network and making a momentum superposition. For topological models, these excitations additionally have an MPO string attached, one or more of the $O_a$, which implements their topological character, such as braiding and spin. Due to the pulling through condition (see Figure \ref{PullThrough}), these strings are locally undetectable, hence in the topological phase no energy penalty is associated with them. For MPO-injective PEPS, it was shown in \cite{bultinck2015anyons} that for an excitation to be of a single well-defined anyon type, the locally modified tensor has to live exactly in the support of an idempotent, one for every anyon type. These idempotents are uniquely defined by and can be obtained from the MPOs $O_a$ that characterize the PEPS, see Figure \ref{fig:Excit}. In the case of abelian quantum doubles such as the Toric Code, these idempotents would exactly correspond to the more familiar projectors on the different charge and flux sectors.

\begin{figure}[h!]
\begin{align*} 
\vcenter{\hbox{
\includegraphics[width=0.25\linewidth]{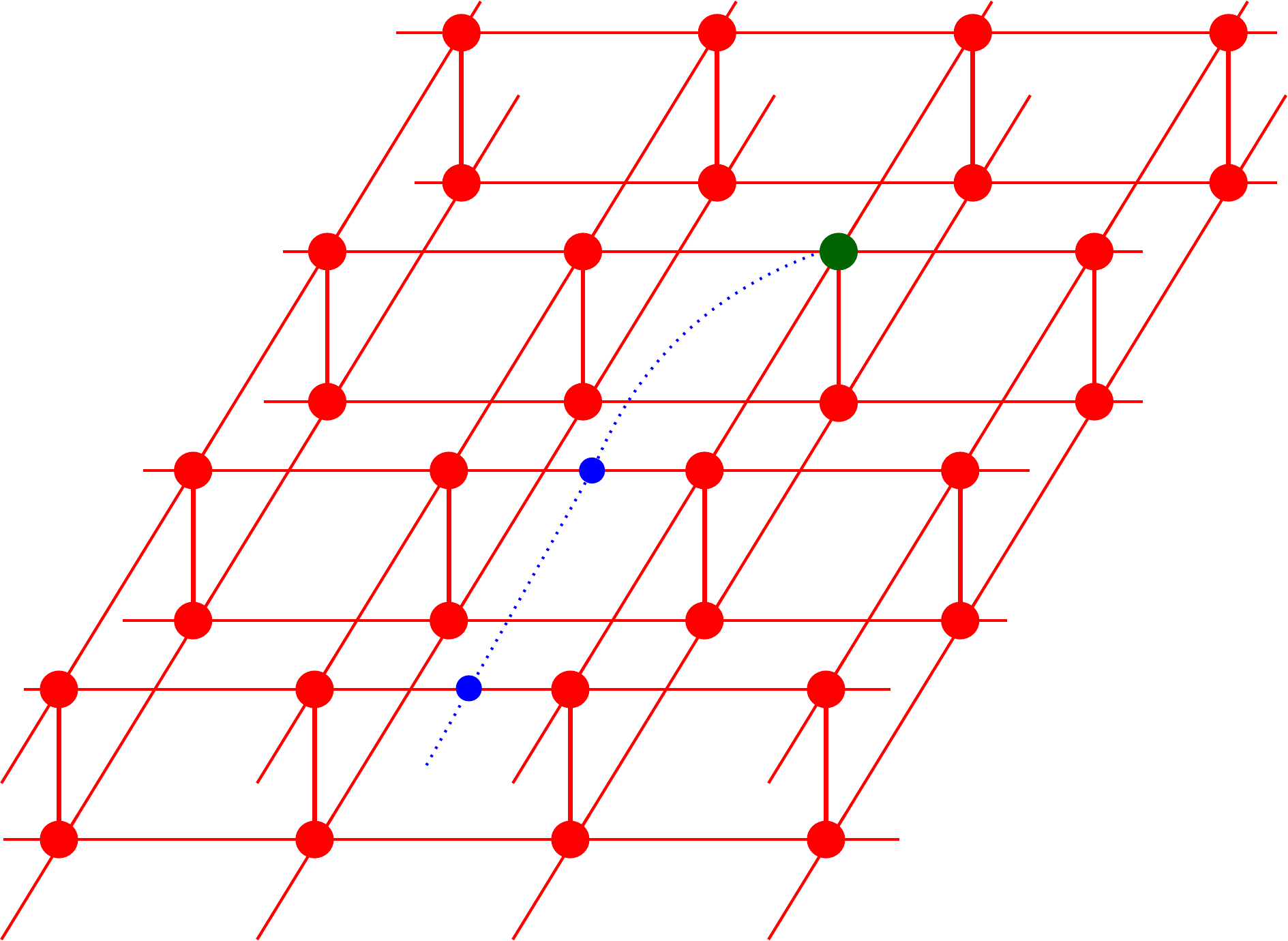}}} = 0, \qquad 
\vcenter{\hbox{
\includegraphics[width=0.25\linewidth]{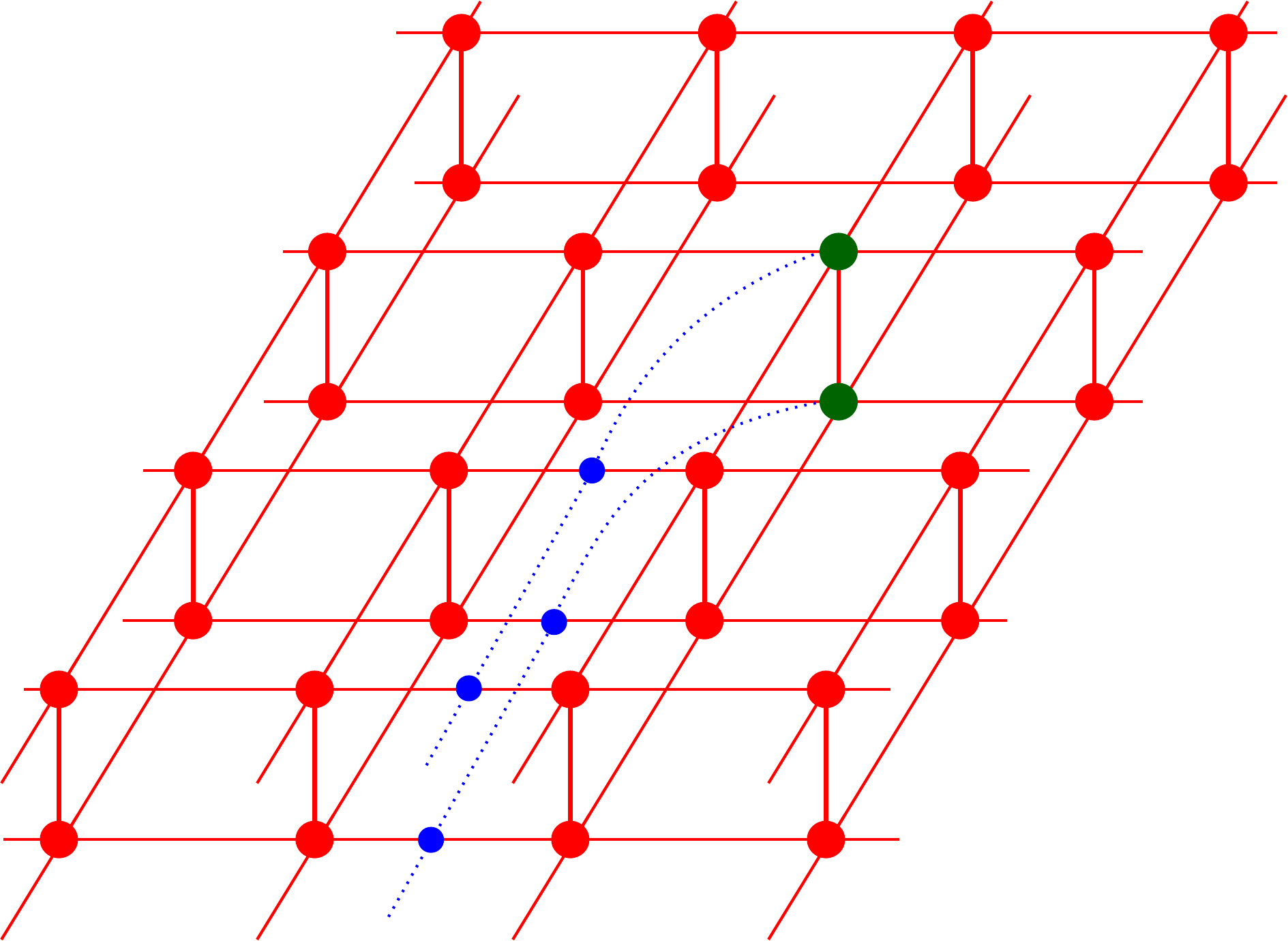}}}\neq 0
\end{align*} 
\caption{The conditions to have a well defined, particle like, anyon excitation. The left condition expresses that the excitation is not condensed, the right one that it is not confined.}
\label{fig:NormEx}
\end{figure}

Let us now denote the PEPS state determined by the tensor in Figure \ref{fig:MPOinvariant} as $\ket{\psi}$ and the state with an extra excitation as in Figure \ref{fig:Excit} as $\ket{\psi[O]}$. For the excitation to be well defined we need it to be orthogonal to the ground state and have a well-defined, non-vanishing norm. These conditions are given by $\braket{\psi|\psi[O]} = 0$ and $\braket{\psi[O]|\psi[O]} \neq 0$ and illustrated in Figure \ref{fig:NormEx}. As we explain below, we can interpret a violation of these conditions in terms of particle confinement or condensation: if $\braket{\psi|\psi[O]} \neq 0$, the particle is condensed, whereas if  $\braket{\psi[O]|\psi[O]} = 0$, the particle is confined.

\begin{figure}[h!]
\begin{align*}
\braket{\psi|\psi}=
\vcenter{\hbox{
\includegraphics[height=0.2\linewidth]{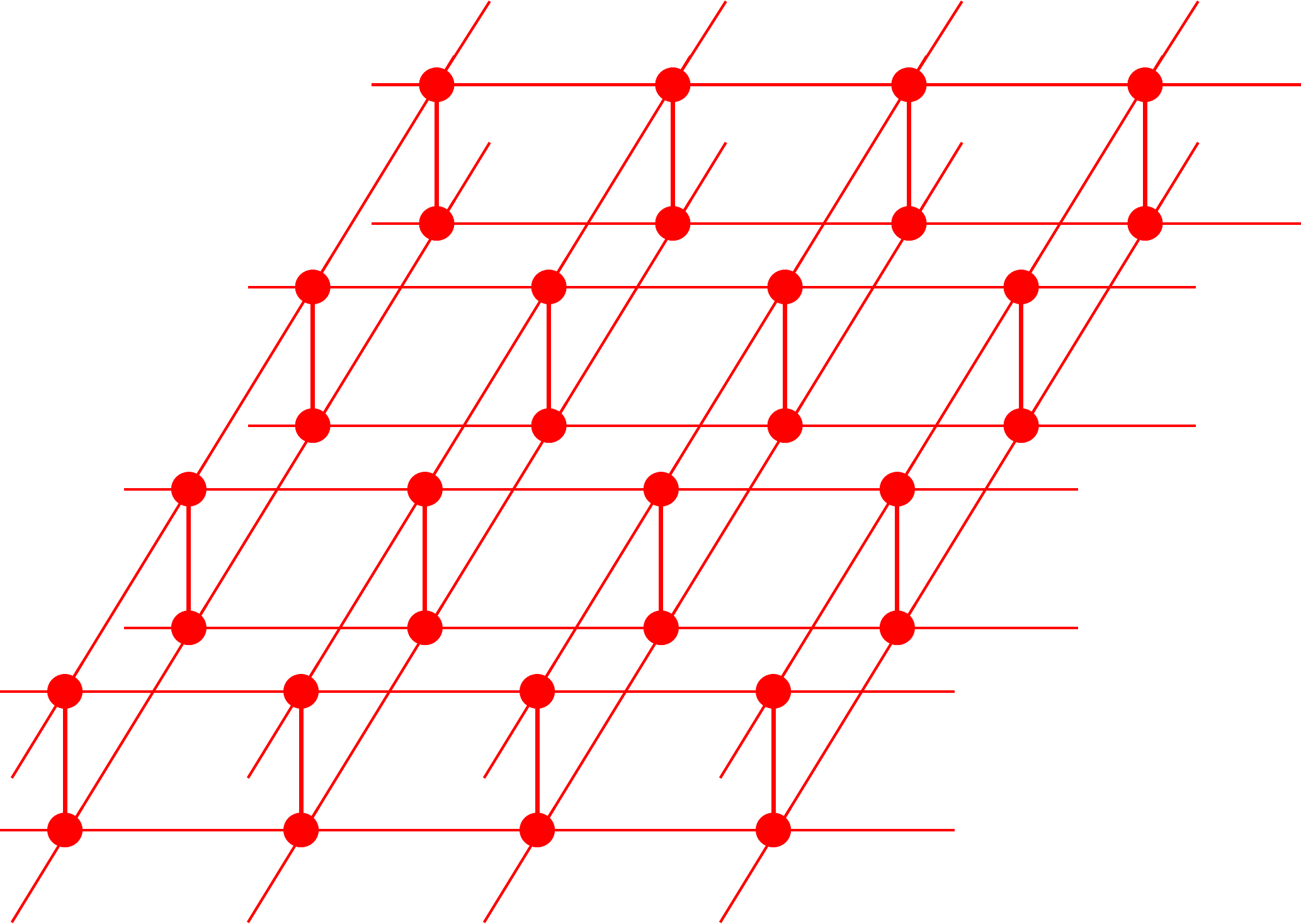}}} 
\quad
\vcenter{\hbox{
\includegraphics[height=0.2\linewidth]{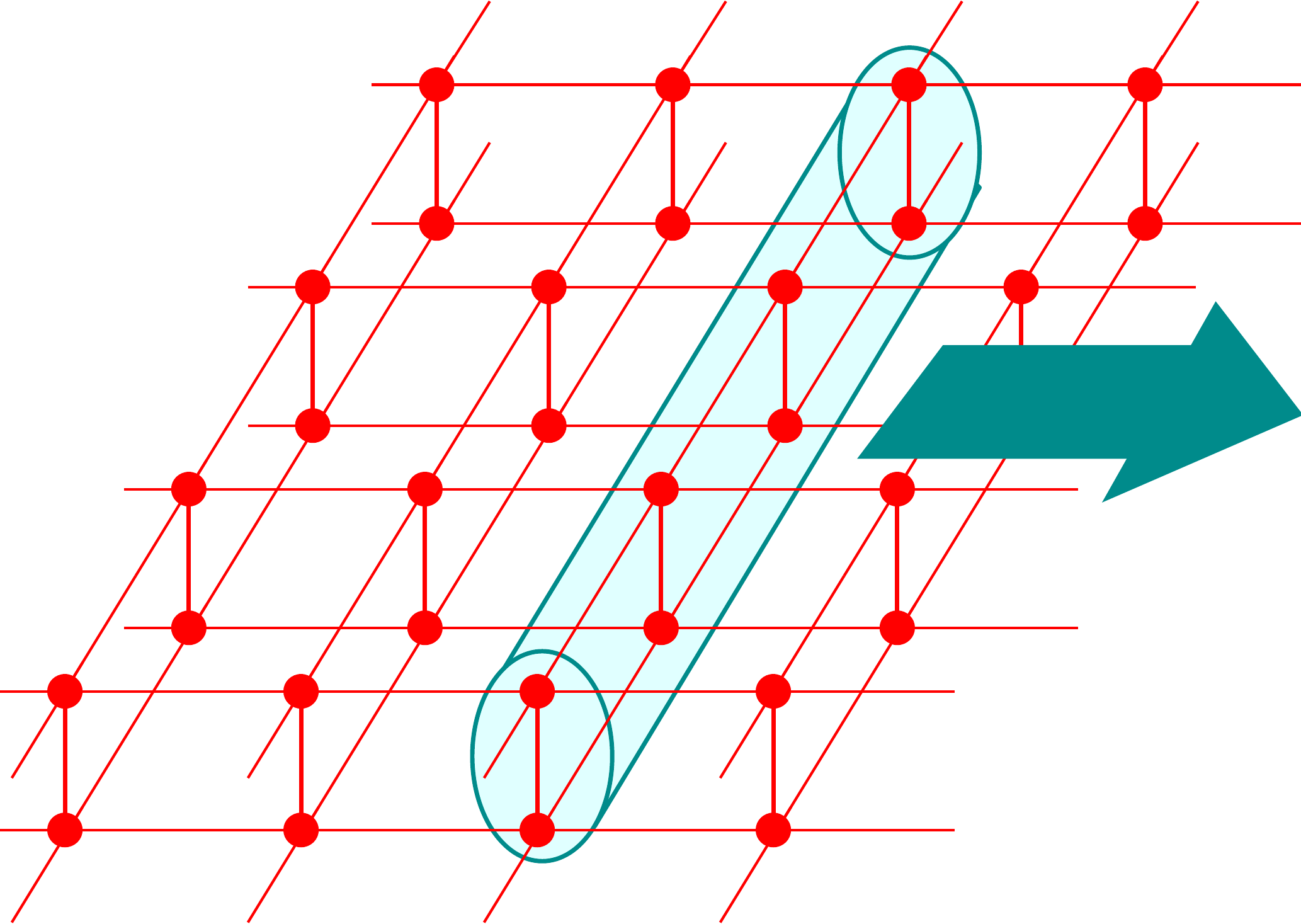}}} 
\vcenter{\hbox{
\includegraphics[height=0.2\linewidth]{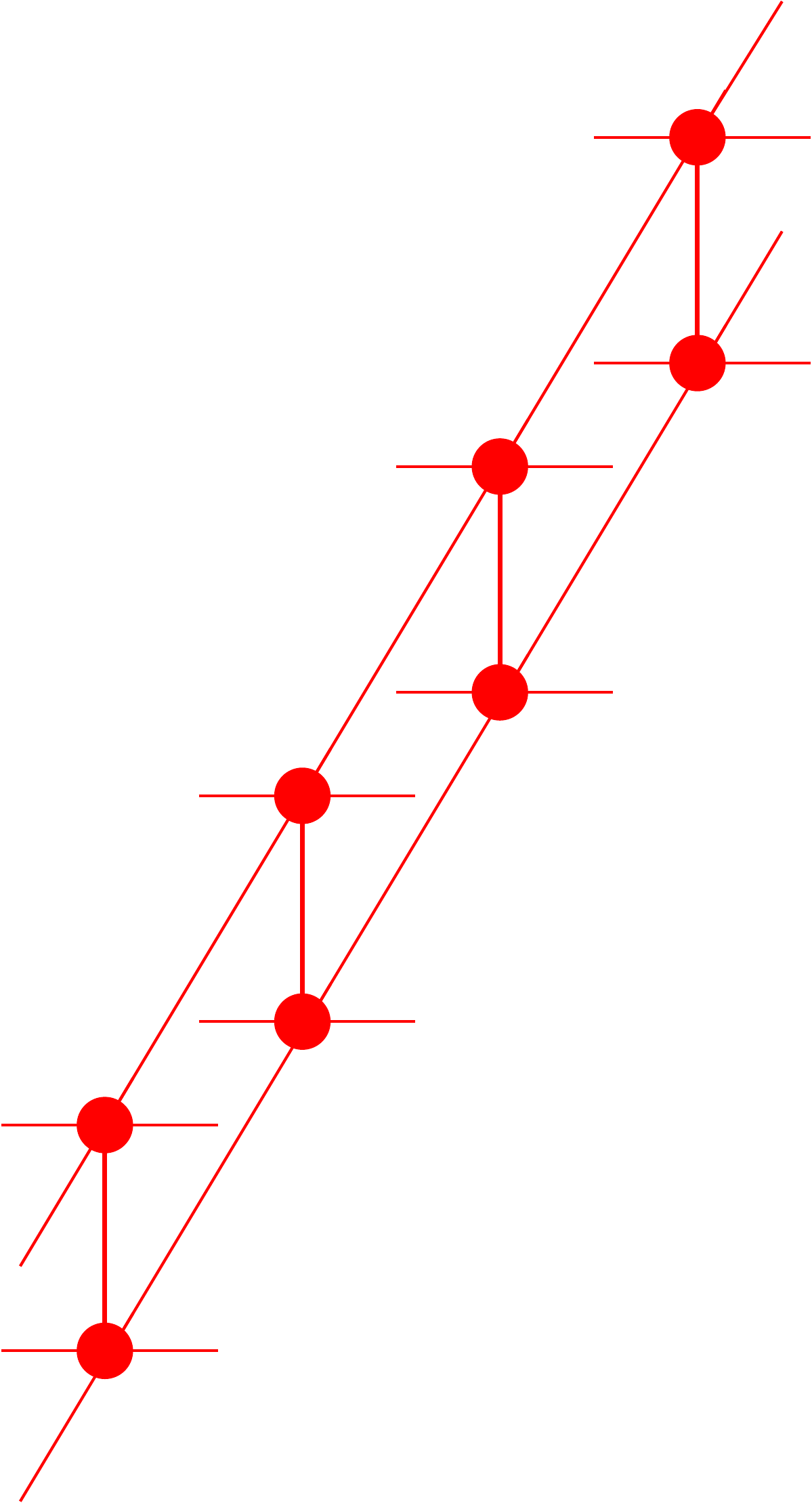}}} 
\end{align*}
\caption{The norm of a PEPS is given by the full contraction of the double layer network (left). We can take one column of the double layer and treat it as an operator from left to right, the transfer matrix (right).}
\label{fig:transfermatrix}
\end{figure}

The previous two conditions can be interpreted in terms of the symmetries of the so called transfer matrix. To introduce this operator, we look at the norm of the PEPS $\ket{\psi}$. The norm of a PEPS is given by the value of the double layer contracted network, one column of this network can be seen as an operator and is called the transfer matrix, see Figure \ref{fig:transfermatrix}. We can treat such a column as an operator, the transfer matrix, acting on the open indices on the right. The normalization of the PEPS implies that the largest eigenvalue of this operator is 1. 

\begin{figure}[h!]
\begin{align*}
\vcenter{\hbox{
\includegraphics[height=0.2\linewidth]{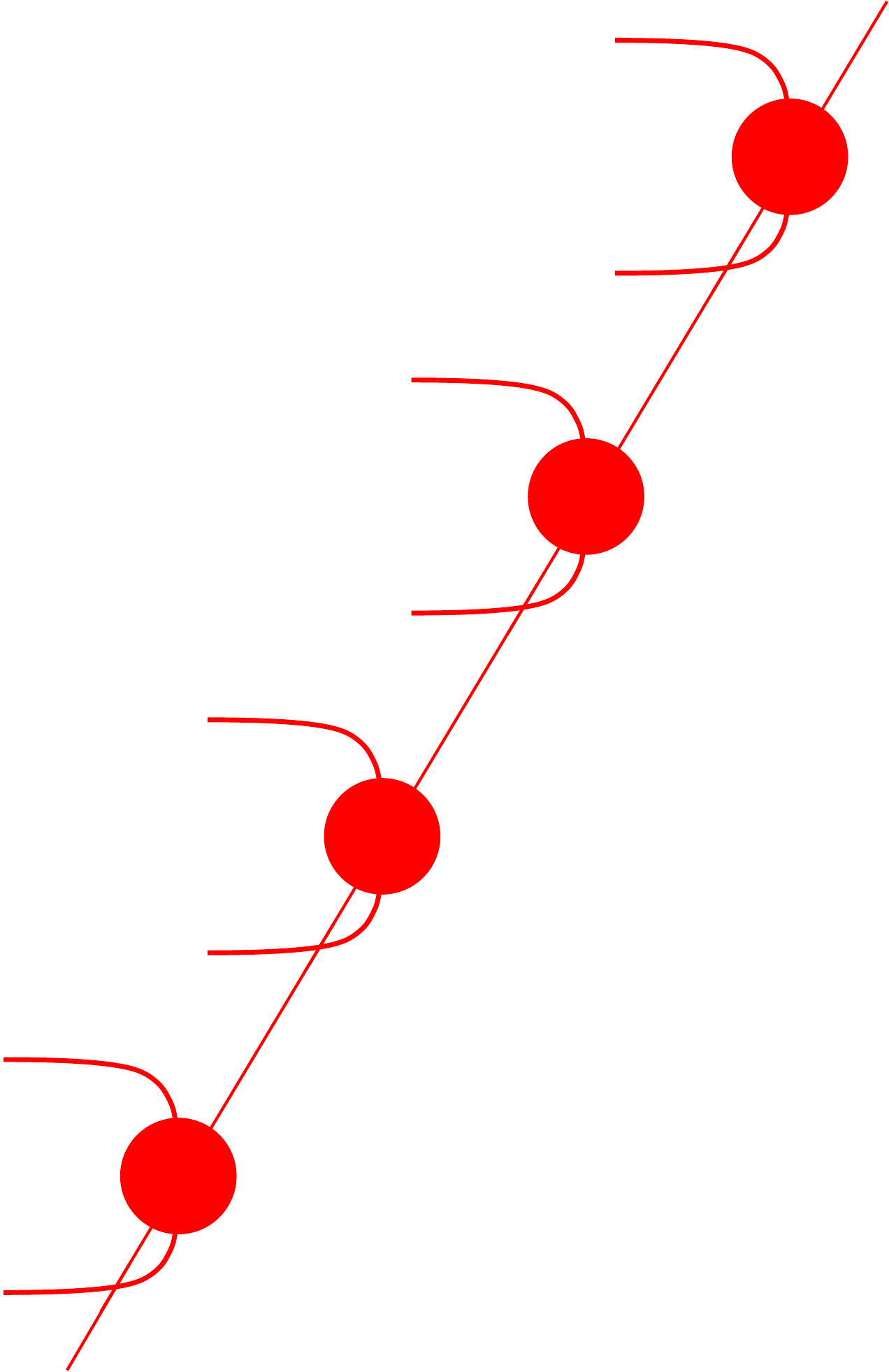}}}, \qquad
\vcenter{\hbox{
\includegraphics[height=0.2\linewidth]{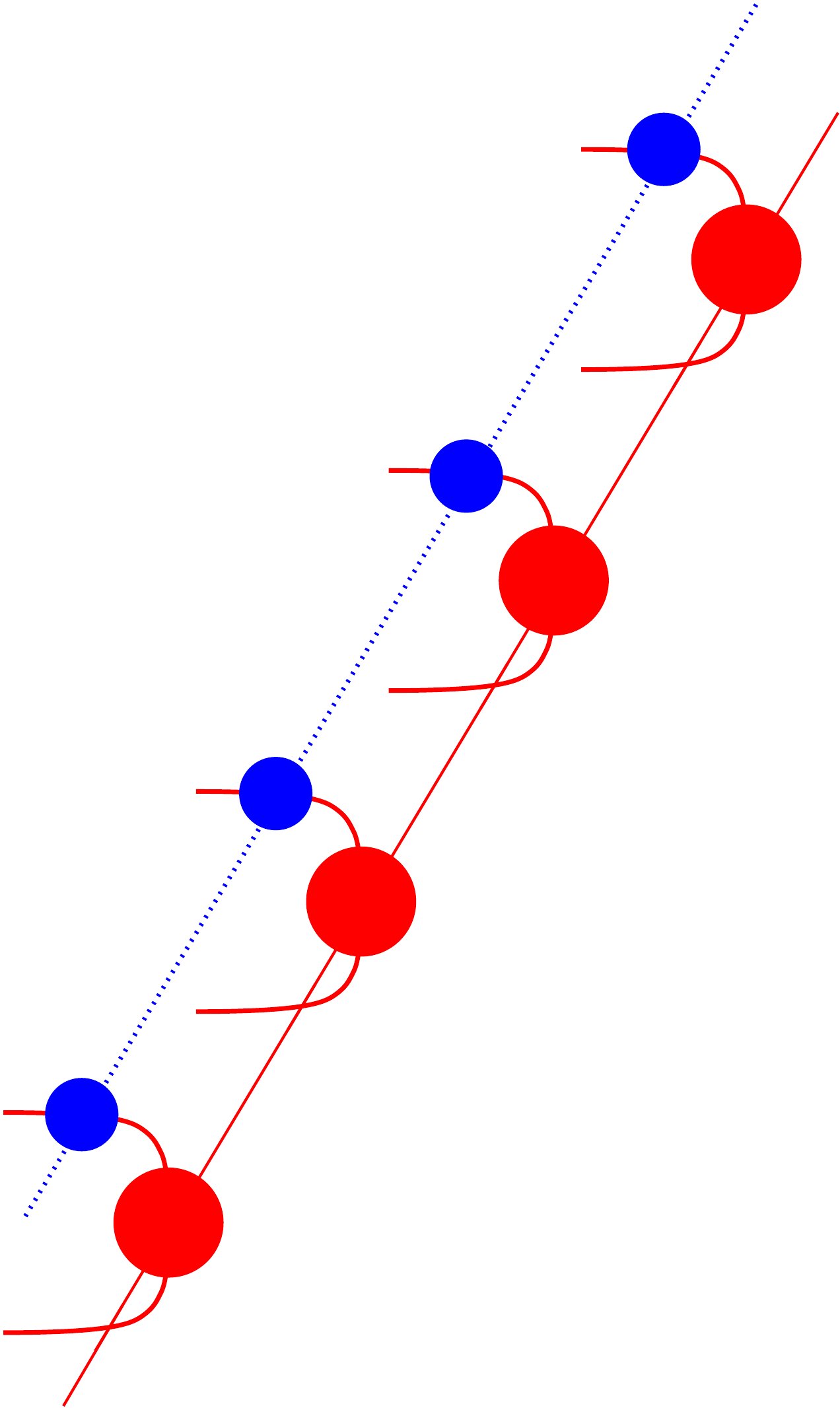}}}=
n_1 \vcenter{\hbox{
\includegraphics[height=0.2\linewidth]{FixedPoint.pdf}}}+
n_2\vcenter{\hbox{
\includegraphics[height=0.2\linewidth]{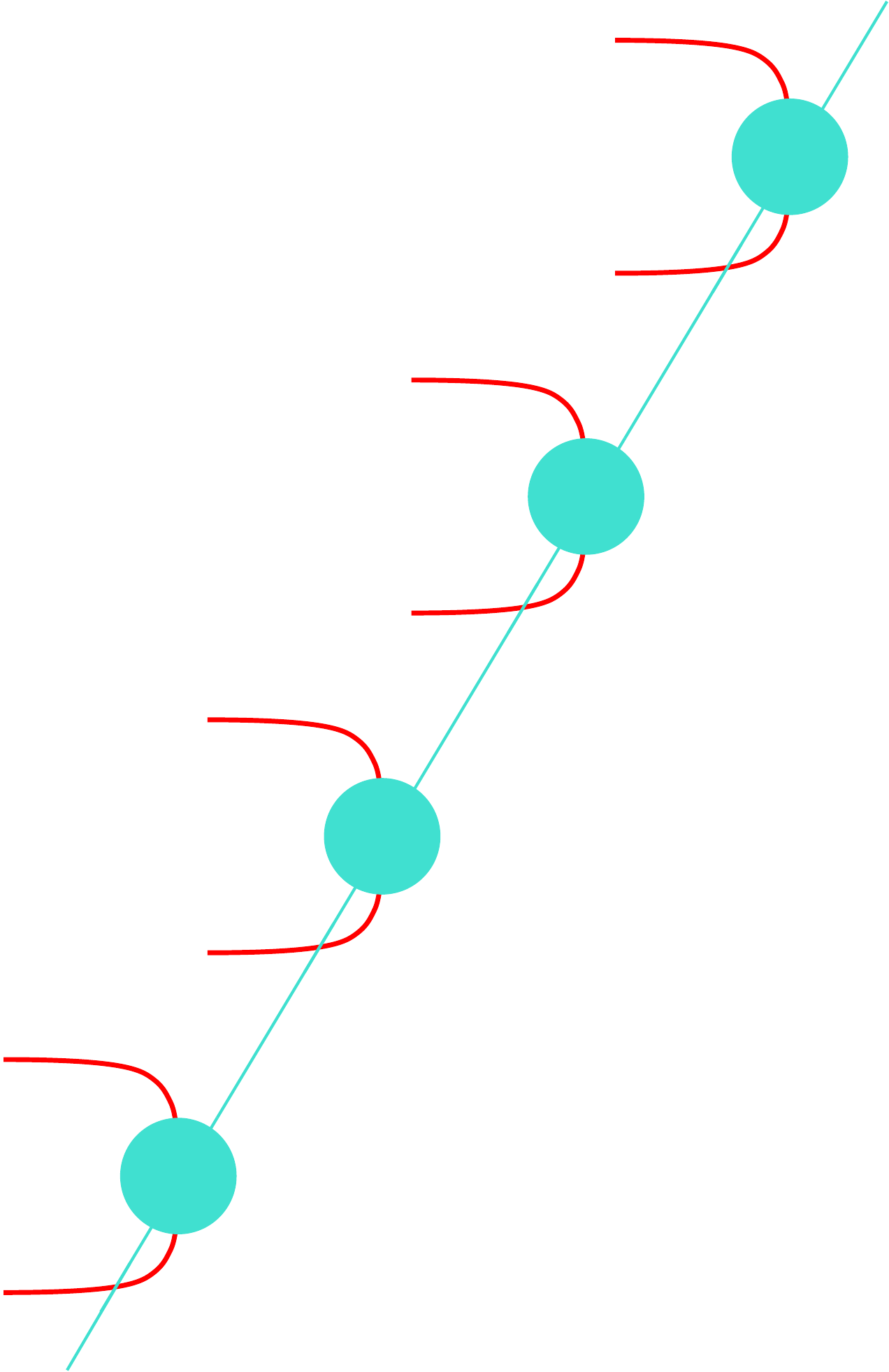}}}+... 
\end{align*}
\caption{A fixed point of the transfer matrix, represented as MPS (left). Applying an MPO to a fixed point (centre) gives us another fixed point, as a possibly non-injective MPS. It can be expanded as a combination of the injective fixed points with integer coefficients $n_i$ (right).}
\label{fig:fixedpoint}
\end{figure}

We denote the, possibly degenerate, fixed points of the transfer matrix as in Figure \ref{fig:fixedpoint}. Because of the pulling though property of the MPOs, given a fixed point we can construct more fixed points by applying a specific MPO in the bra or ket layer. The application of such an MPO results in a new, possibly non injective, fixed point which can be expanded as a sum of injective fixed points. We use different colors to stress that there are different fixed points, or use an extra label. In the RG fixed point, the fixed points are essentially given by the injective MPOs $O_a$. For string nets, they follow a commutative fusion algebra This implies that there is only one possible positive operator among them, which is $O_1$, the MPO that acts as the identity in the PEPS network. We denote the topologically trivial fixed point obtained from this MPO as $\rho_1$. 

\begin{figure}[h!]
\begin{align*}
\braket{\psi[O] |\psi[O]}=
\vcenter{\hbox{
\includegraphics[height=0.2\linewidth]{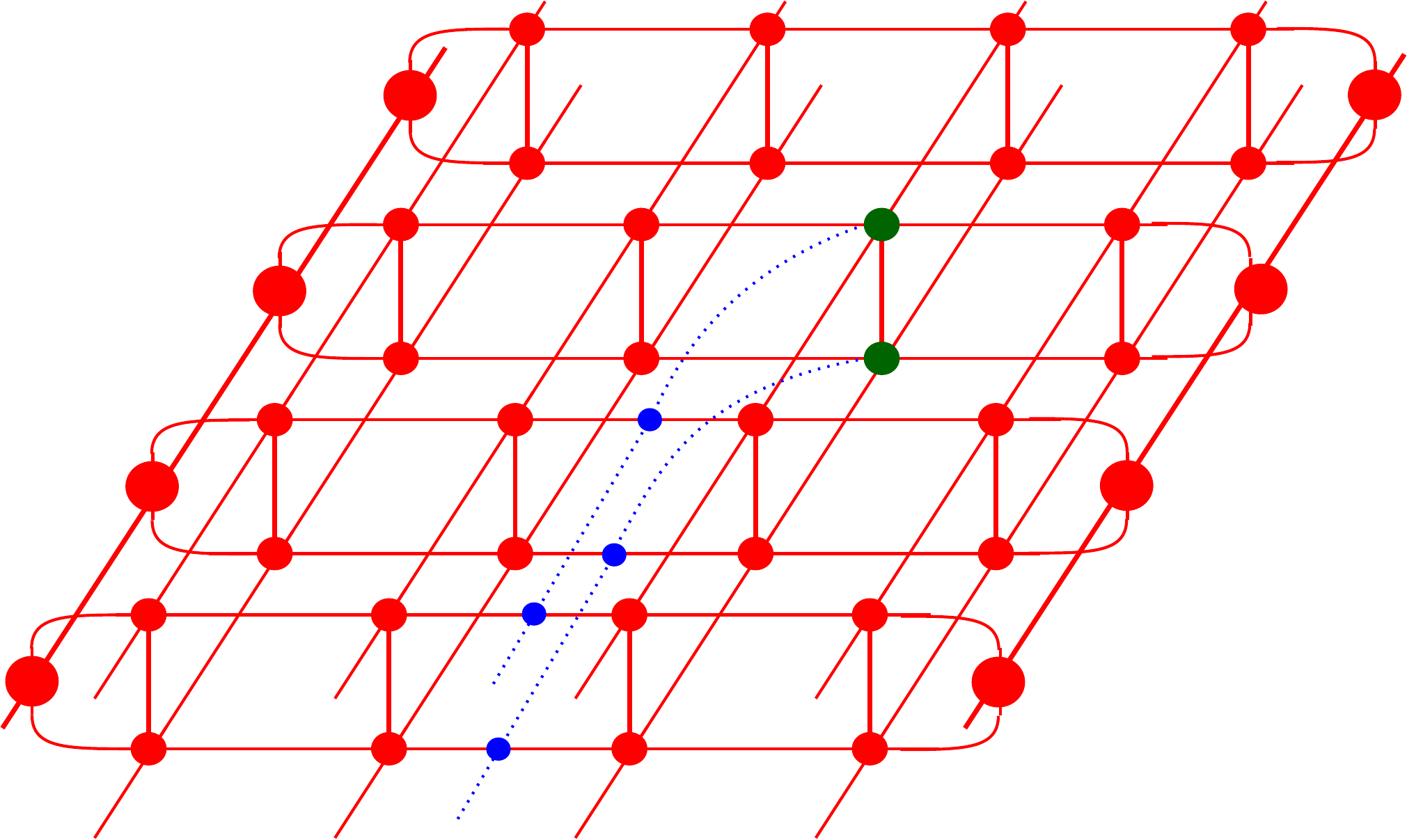}}}=
\vcenter{\hbox{
\includegraphics[height=0.2\linewidth]{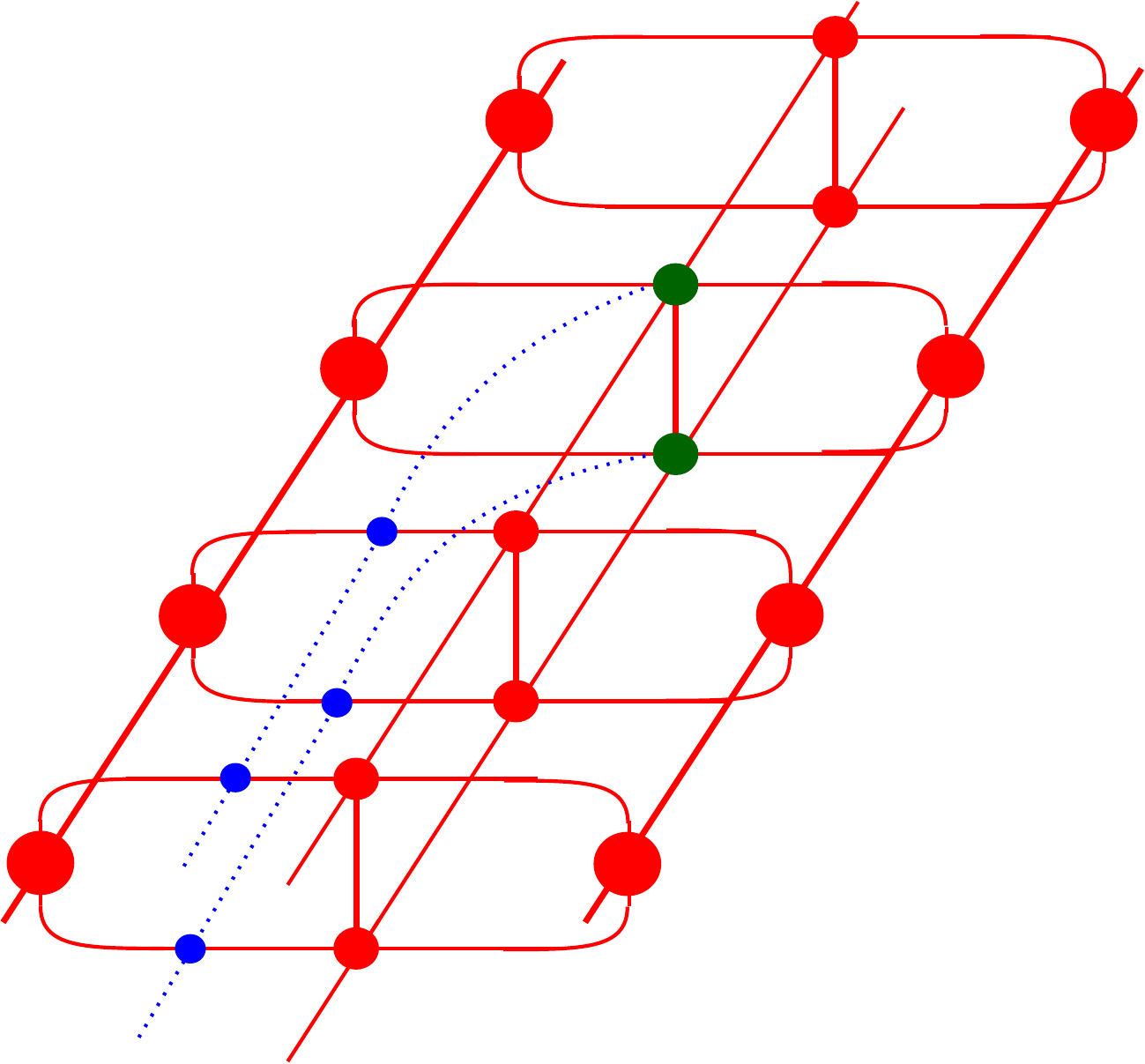}}}=
\vcenter{\hbox{
\includegraphics[height=0.2\linewidth]{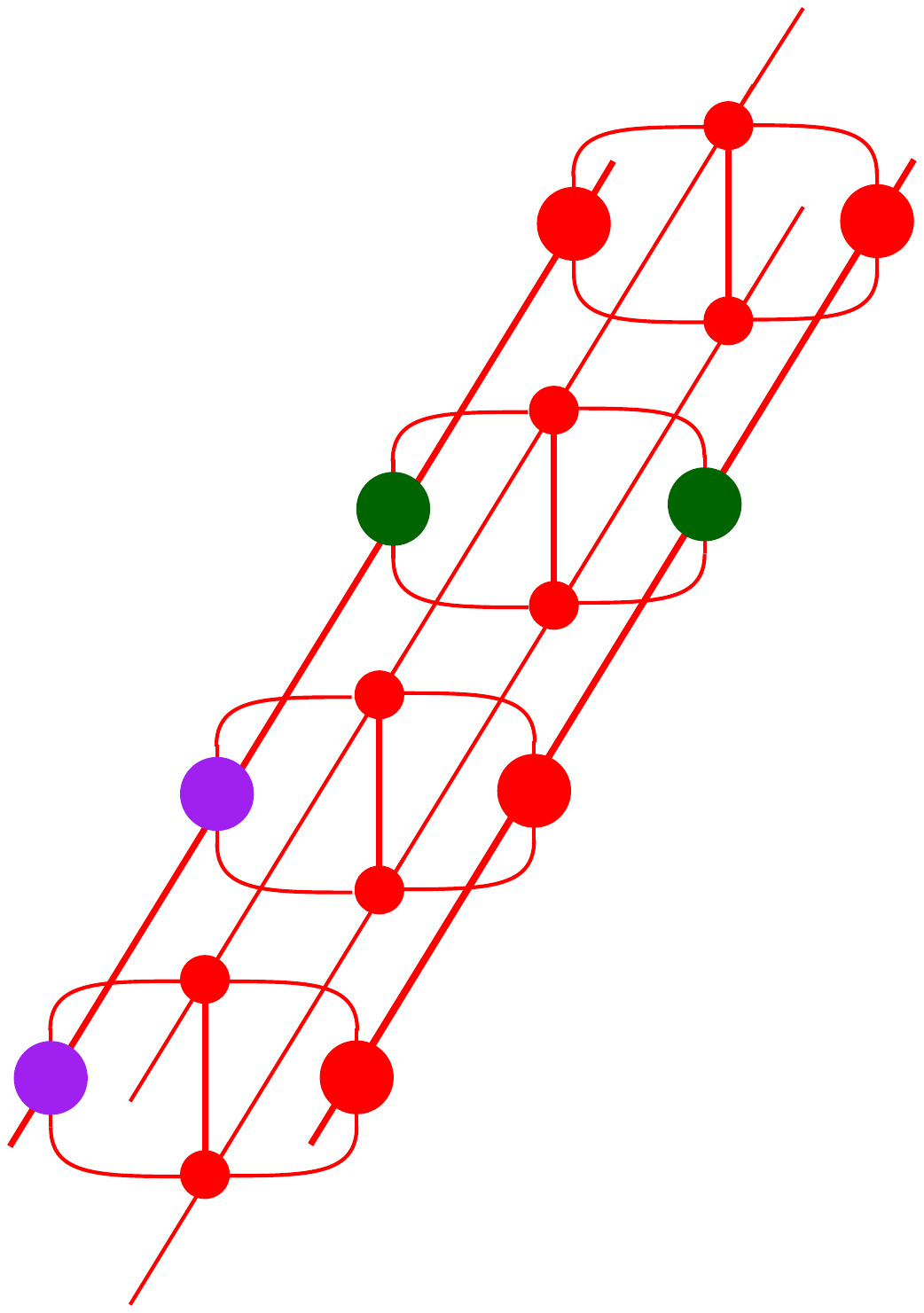}}}
\end{align*}
\caption{Contraction scheme to calculate the norm of an excitation, see the main text for details.}\label{fig:normex2}
\label{fig:normex}
\end{figure}

\begin{figure}[h!]
\begin{align*}
\braket{\psi |\psi[O]}=
\vcenter{\hbox{
\includegraphics[height=0.2\linewidth]{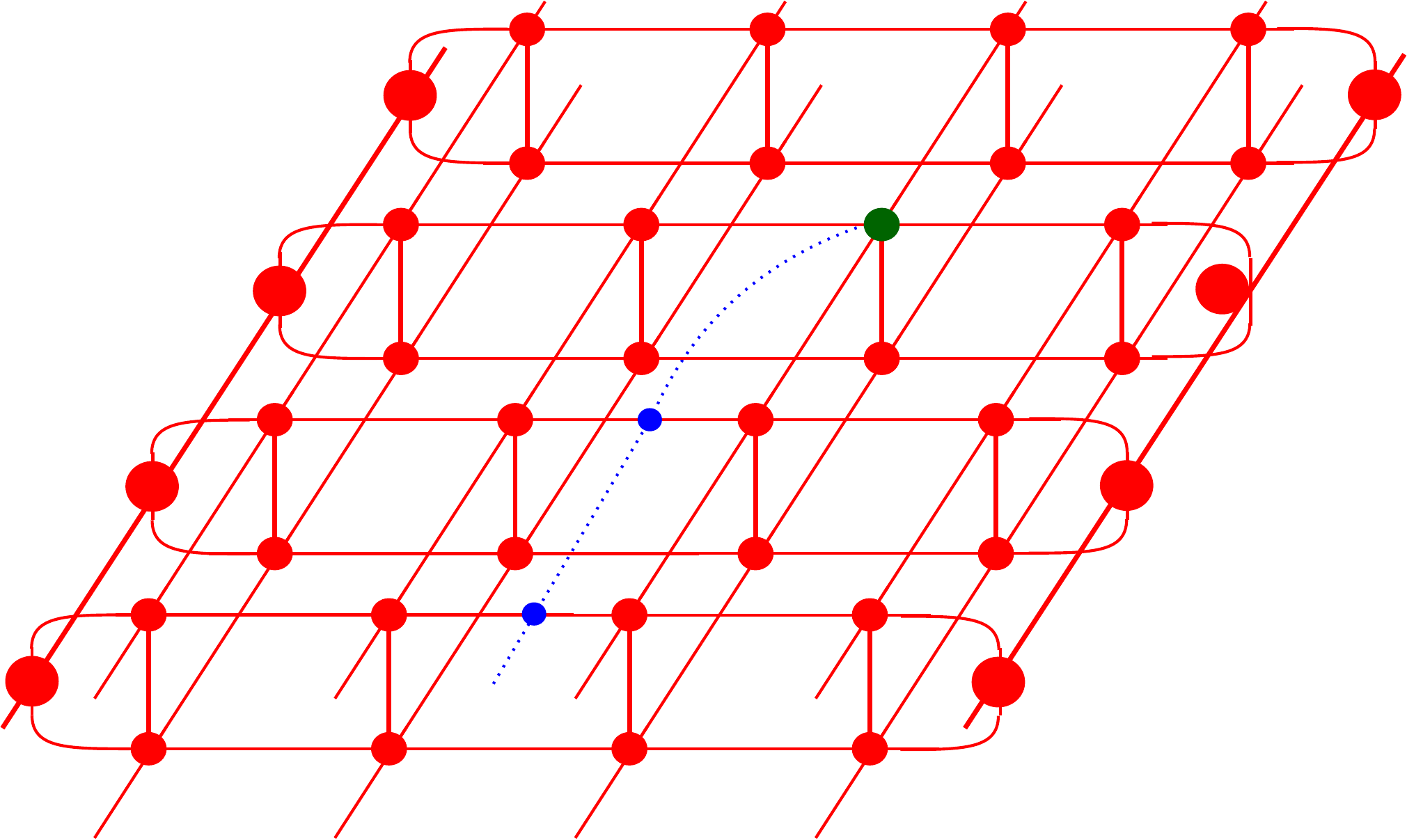}}}=
\vcenter{\hbox{
\includegraphics[height=0.2\linewidth]{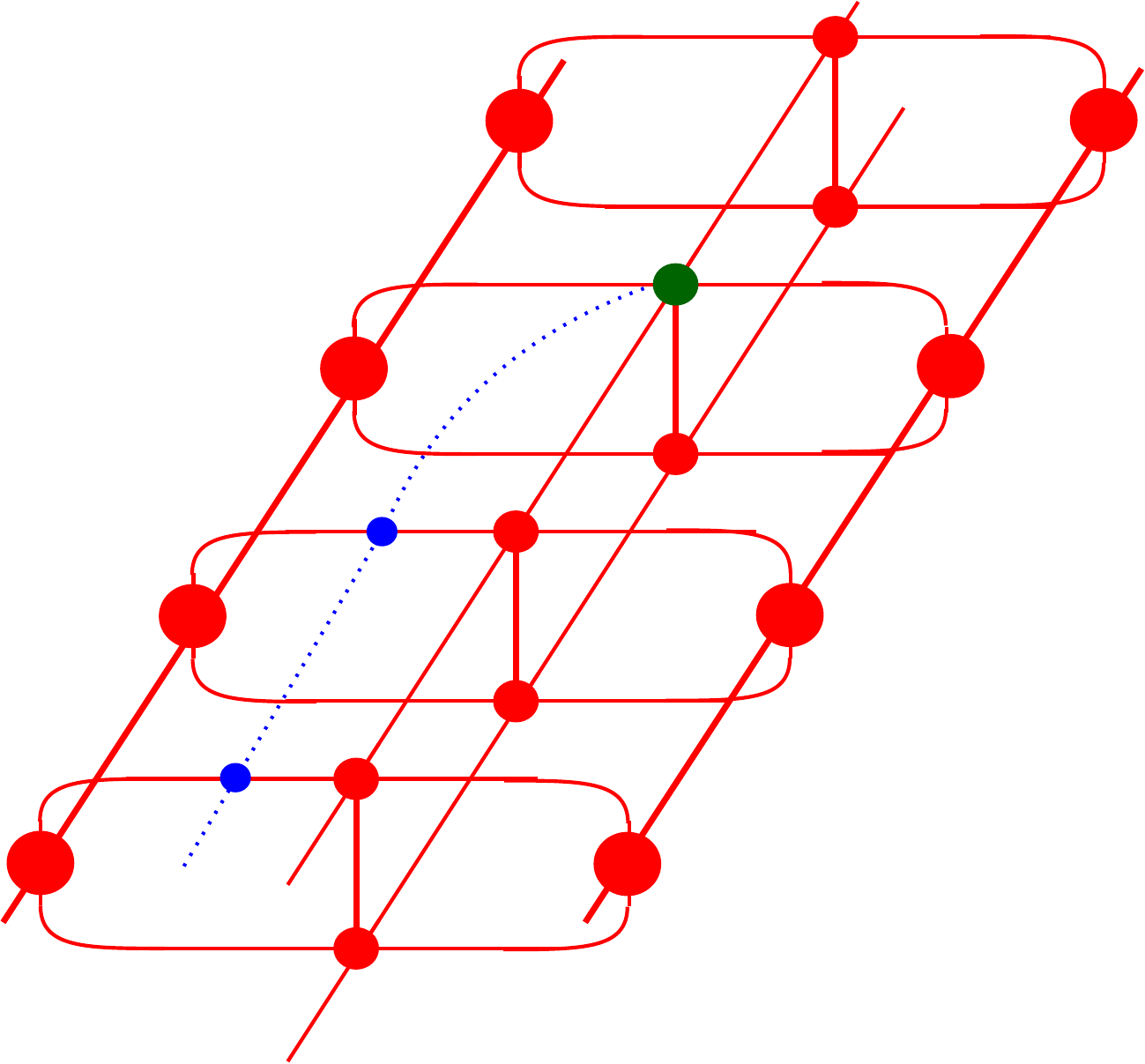}}}=
\vcenter{\hbox{
\includegraphics[height=0.2\linewidth]{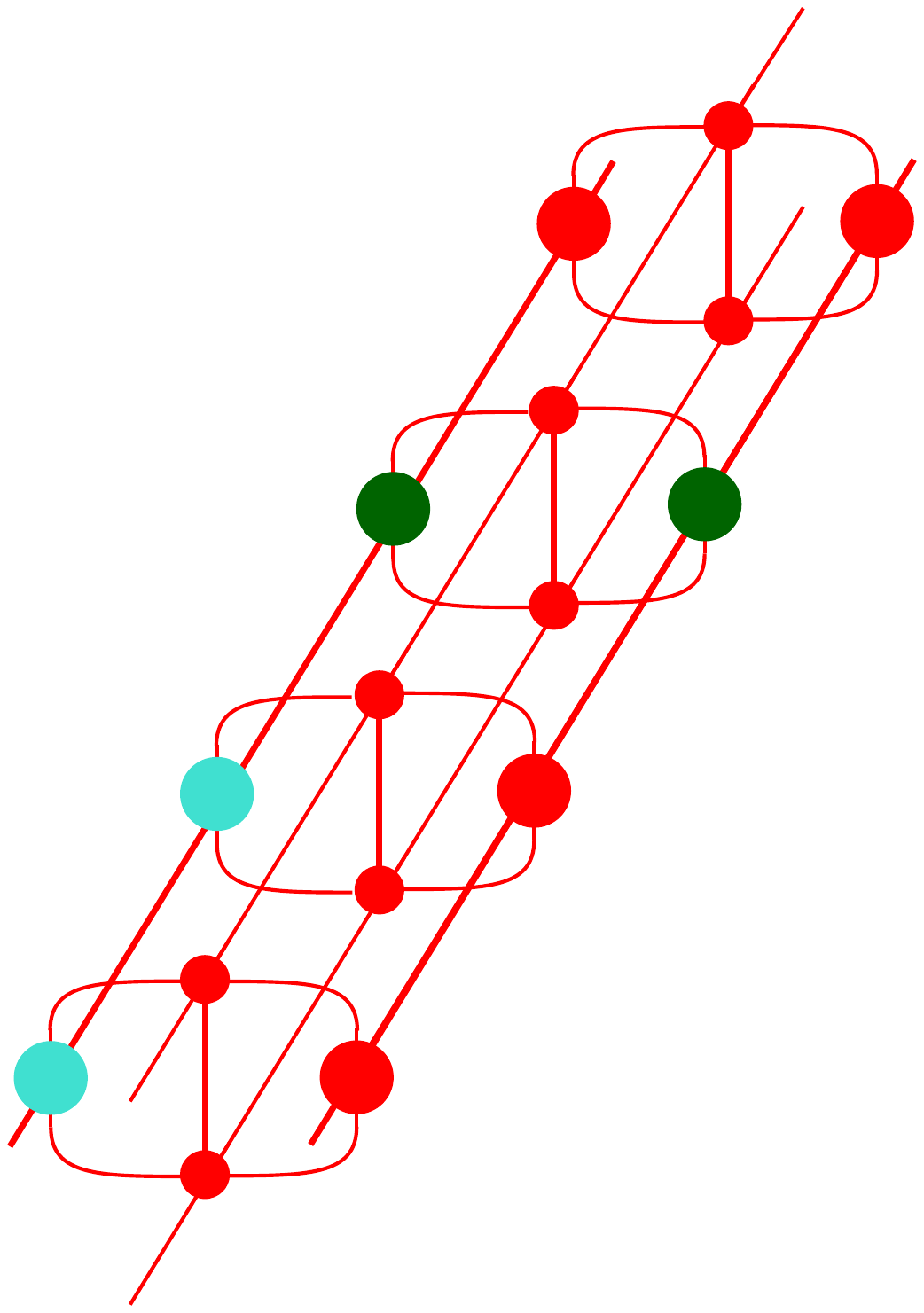}}}
\end{align*}
\caption{Contraction scheme to calculate the overlap of the ground state and an excited state, see the main text for details.}
\label{fig:overlap}
\end{figure}

We now return to the conditions in Figure \ref{fig:NormEx}. To compute the norm of $\ket{\psi[O]}$ we proceed as follows. The contraction of the entire network to the left or right of the position of the excitation tensor gives us the unique positive, left and right fixed point of the transfer matrix as illustrated in Figure \ref{fig:normex}. Similarly, the overlap between the ground state $\ket{\psi}$ and the excited state $\ket{\psi[O]}$ is given in Figure \ref{fig:overlap}.

\begin{figure}[h!]
\begin{equation*}
\lambda_{\text{max}}\Bigg(
\vcenter{\hbox{
\includegraphics[height=0.08\linewidth]{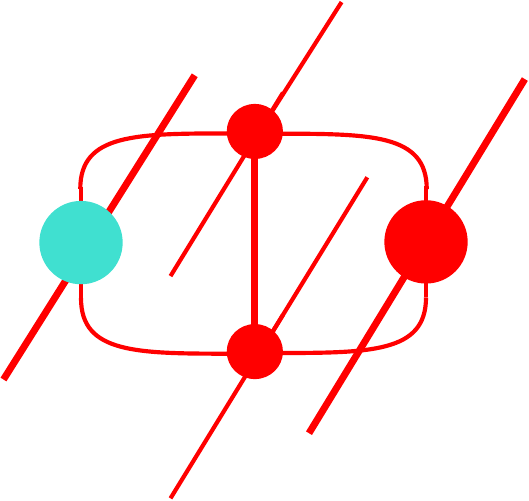}}} \: \Bigg) <1, \qquad 
\lambda_{\text{max}}\Bigg(
\vcenter{\hbox{
\includegraphics[height=0.08\linewidth]{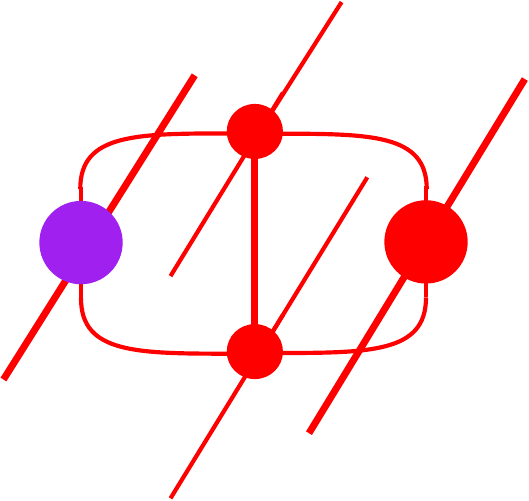}}} \: \Bigg) =1 
\end{equation*}
\caption{Conditions necessary to have a well-defined excitation with blue MPO string. The figures on left and right are to be interpreted as matrices form the upper to the lower indices.}
\label{fig:overlapmat}
\end{figure}

We see that in order to have well defined excitation with an MPO string, i.e.\ $\braket{\psi|\psi[O]}=0$ and $\braket{\psi[O]|\psi[O]} \neq 0$ as depicted in \ref{fig:NormEx}, it is sufficient if the conditions in Figure \ref{fig:overlapmat} are fulfilled. The second condition (on the right of Figure \ref{fig:overlapmat}) is also necessary for having $\braket{\psi[O]|\psi[O]} \neq 0$, whereas $\braket{\psi|\psi[O]}=0$ could still be satisfied thanks to the local tensor at the endpoint of the string excitation even if the left condition of Figure \ref{fig:overlapmat} is not satisfied. One can think of the strings as the flux degrees of freedom and the endpoint as the charge degrees of freedom, although the two can not always be separated in such an easy manner. Where the MPOs implement the symmetry action on the transfer matrix, the endpoints can be thought of as order parameters for these symmetries. This correspondence was further investigated in \cite{haegeman2015shadows} for phases corresponding to doubled group algebras (and in particular the Toric Code phase). In our more general setting, the idempotents living at the end point of the string are more complex, and this interpretation as virtual order parameter is beyond the current scope. Nevertheless, it is clear that the conditions in Figure \ref{fig:overlapmat} provide important information about the structure of the fixed point subspace. The transfer matrix commutes with the application of all MPOs, both on the bra and the ket level, as follows from the pulling through condition in Figure \ref{PullThrough}. Given this symmetry, the question remains what the corresponding structure of the fixed point subspace of the transfer matrix is with respect to this symmetry. The left condition in Figure \ref{fig:overlapmat} expresses the fact that the fixed point subspace is not symmetric under the application of an MPO in only the ket (or only the bra) layer. The right condition on the other hand tells us that the fixed point subspace has to be symmetric under the application of the same MPO in both bra and ket layer. 

For the Fibonacci model, we need exactly two fixed points to have well defined anyons. From the fact that the transfer matrix commutes with the application of $O_1,O_{\tau}$ in its bra and ket layer, we expect four different fixed points. Suppose all the candidates are indeed different and consider an excitation with a $O_{\tau}$ MPO string. To calculate its norm we look at the situation described in Figures \ref{fig:normex2} and the right hand side of \ref{fig:overlapmat} from which we conclude that anyons with such a string  have a zero norm. This means they cannot appear on their own and are thus confined. On the other hand, if all fixed point candidates are equal, the overlap between a state with such an excitation and the ground state is described in Figures \ref{fig:overlap} an the left hand side of \ref{fig:overlapmat}. We see that this overlap is nonzero and conclude that if all fixed points are equal, the excitations are condensed. To have well defined anyons with a $O_{\tau}$ string we thus demand exactly two fixed points. This corresponds to the fact that the fixed point subspace is not invariant under the application of a $O_{\tau}$ MPO in only the ket or bra layer but is invariant under a simultaneous application in both.

\subsection{The Transfer Matrix as a Probe for Excitations}
Unfortunately, a full variational approach to perturbed string net Hamiltonians is numerically out of reach despite recent progress \cite{phien2015infinite,corboz2016variational,laurens2016variational}. Luckily, we can gain lots of insight by using holographic dimensional reduction, which is very natural in the language of tensor networks. One can argue that most of the information about the dispersion of a Hamiltonian is already encoded in the ground state. By looking at the transfer matrix of the ground state, we can extract this information. As the transfer matrix of a 2D ground state is itself a 1D system, we have reduced the spatial dimension of the problem and can now apply well established numerical methods. As discussed, the symmetries of the fixed point subspace of the transfer matrix already contain a lot of information about the condensation or confinement of anyon excitations.

In a translation invariant system, excitations are naturally described as momentum eigenstates. We now explain why the transfer matrix also contains information about the dispersion relation of such excitations  \cite{zauner2015transfer}. The reason the transfer matrix is so relevant is that the variational dispersion relation of excitations in the PEPS picture is mainly determined by the normalization of the excited states.
Indeed, let us write the ground state of an Hamiltonian $H$ as $\ket{\psi_0}$. As shown in \cite{haegeman2013elementary} we can construct elementary excitations by acting with local operators on the ground state and giving them a specified momentum $k=(k_x,k_y)$. 
Take a local Hermitian operator $O$ with zero ground state expectation value and denote its translates by $O_{xy}$, i.e. $O_{xy}$ acts on the sites centered around $(x,y)$.
Denote the excitation with momentum $k$ created by $O$ with
$\ket{\psi[O,k]} = \sum_{x,y} \exp(-i(k_xx+k_yy)) O_{xy}\ket{\psi_0}.$ 
We can then write the variational expression for the energy of a state with momentum $k$ as
$$E= \min_O\frac{\braket{\psi[O,k]|H|\psi[O,k]}}{\braket{\psi[O,k]|\psi[O,k]}}.$$
We can rewrite the numerator of this last expression as 
$$\braket{\psi_0|\sum_{x,y}e^{i(k_xx+k_yy)}[O_{xy},[H,O_{x'y'}]]\sum_{x',y'}e^{-i(k_xx'+k_yy')|\psi_0}}.$$
Because of the double commutator, this expression diverges at most as $\|O\||\text{supp}(O)|V$ in the system size $V$. In contrast the scaling of he denominator can, for a good choice of $O$ and $k$, be significantly faster, hence the expression for $E$ is mainly determined by the denominator. In particular, excitations with a low variational energy can be obtained by making the denominator as large as possible. For a detailed version of this reasoning, as well as other arguments that support the conclusion, we refer to \cite{zauner2015transfer}.

\begin{figure}[h!]
\begin{align*}
\braket{\psi[O,k] |\psi[O,k]}=
\sum_{x,y}e^{-ik_x\Delta_x-ik_y\Delta}\vcenter{\hbox{
\includegraphics[height=0.25\linewidth]{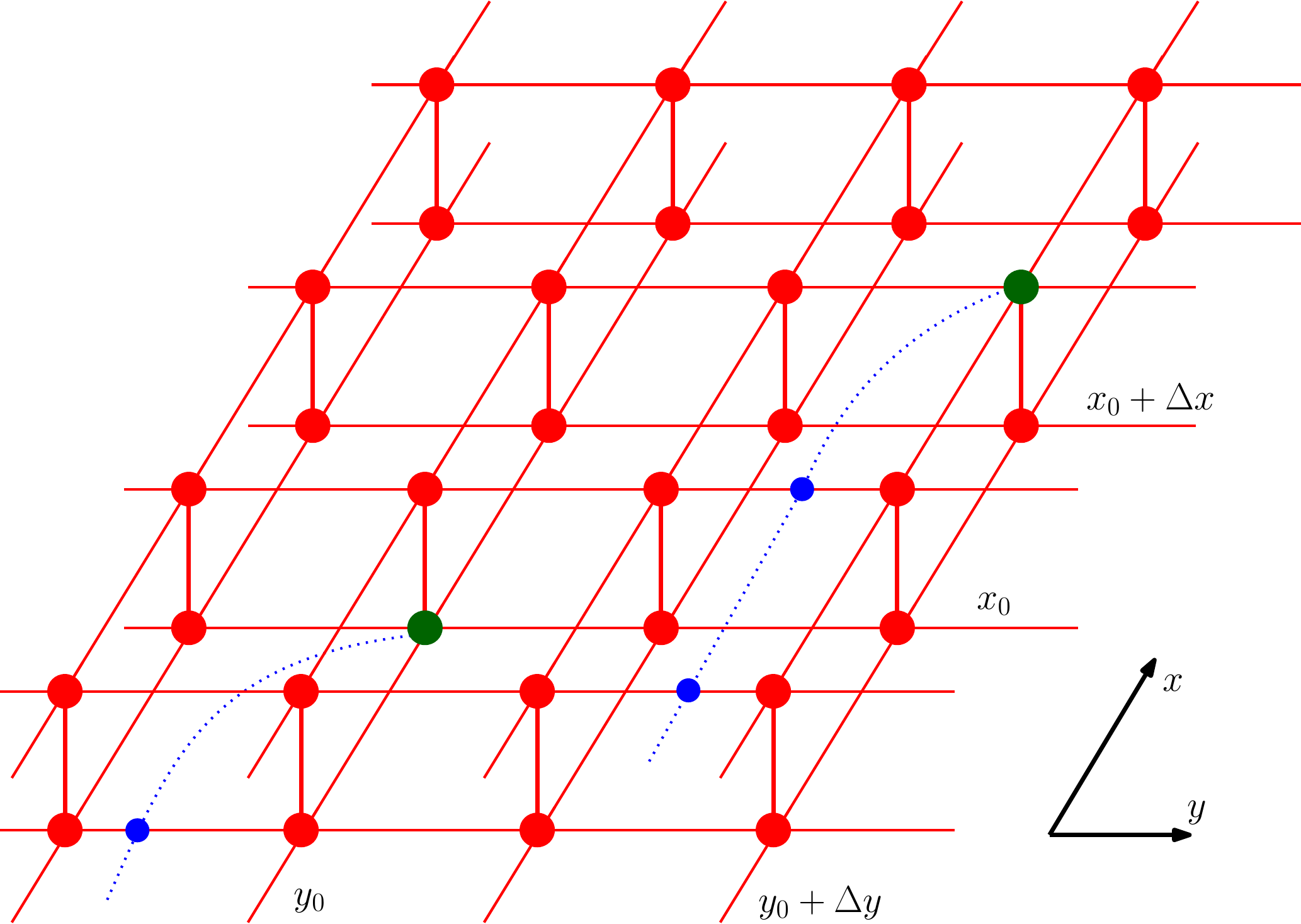}}} 
\end{align*}
\caption{The norm of a excitation, created by acting with a local operator thereby turning a red ground state tensor in a green excited tensor. We make a momentum superposition of this state with momentum $k=(k_x,k_y)$.}
\label{fig:normexcitationPEPS}
\end{figure}

While the previous argument was applicable to point like excitations, it is clear that it can be extended to stringlike excitations of the type above, as the string is unobservable and does not bring along an energy cost. The denominator for such excitations is depicted in Figure \ref{fig:normexcitationPEPS}. It represents the norm of an excited state containing one anyon, localized at the green tensor.
 
\begin{figure}[h!]
\begin{align}
\vcenter{\hbox{
\includegraphics[height=0.2\linewidth]{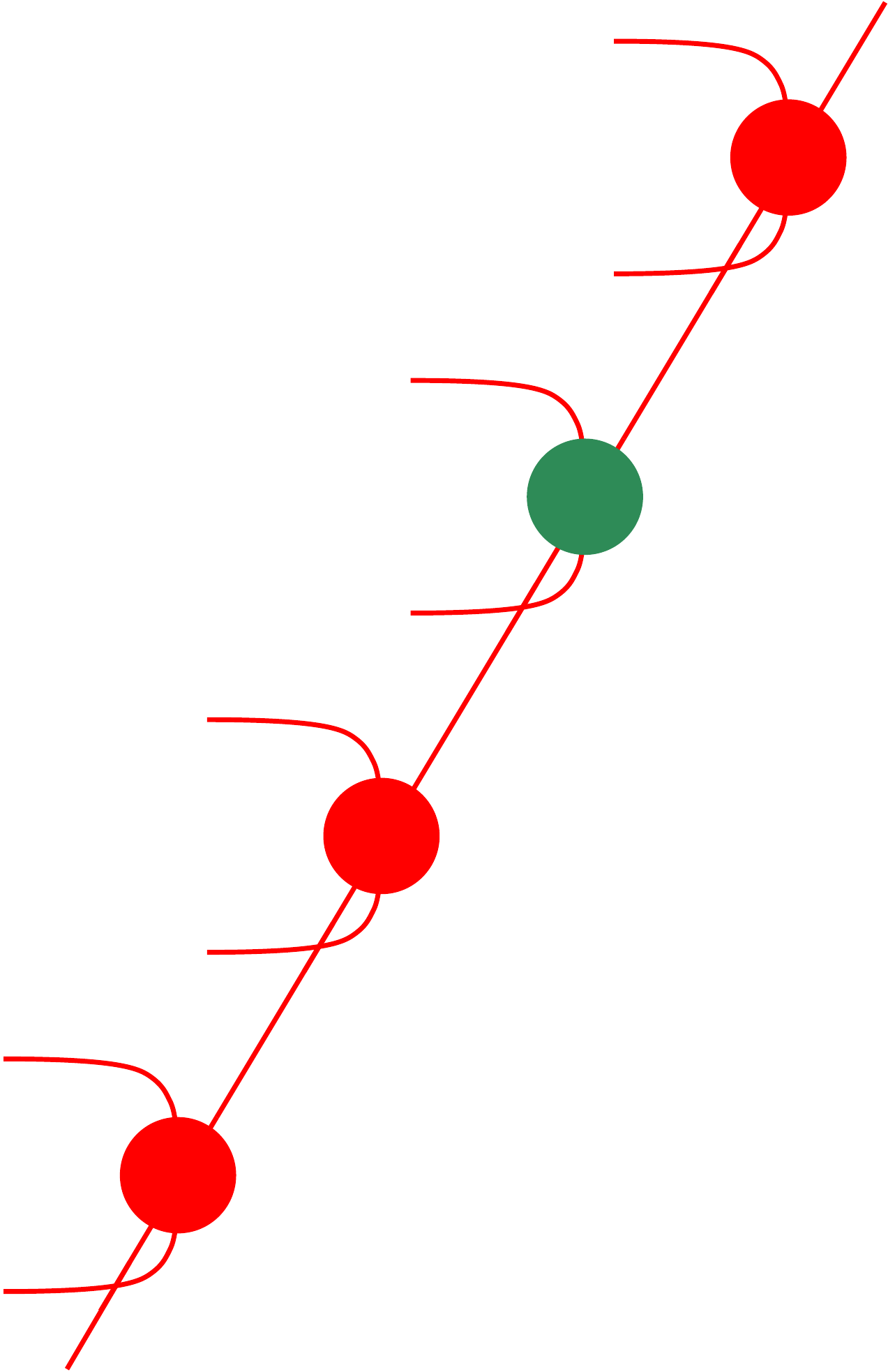}}},\qquad\qquad
\vcenter{\hbox{
\includegraphics[height=0.2\linewidth]{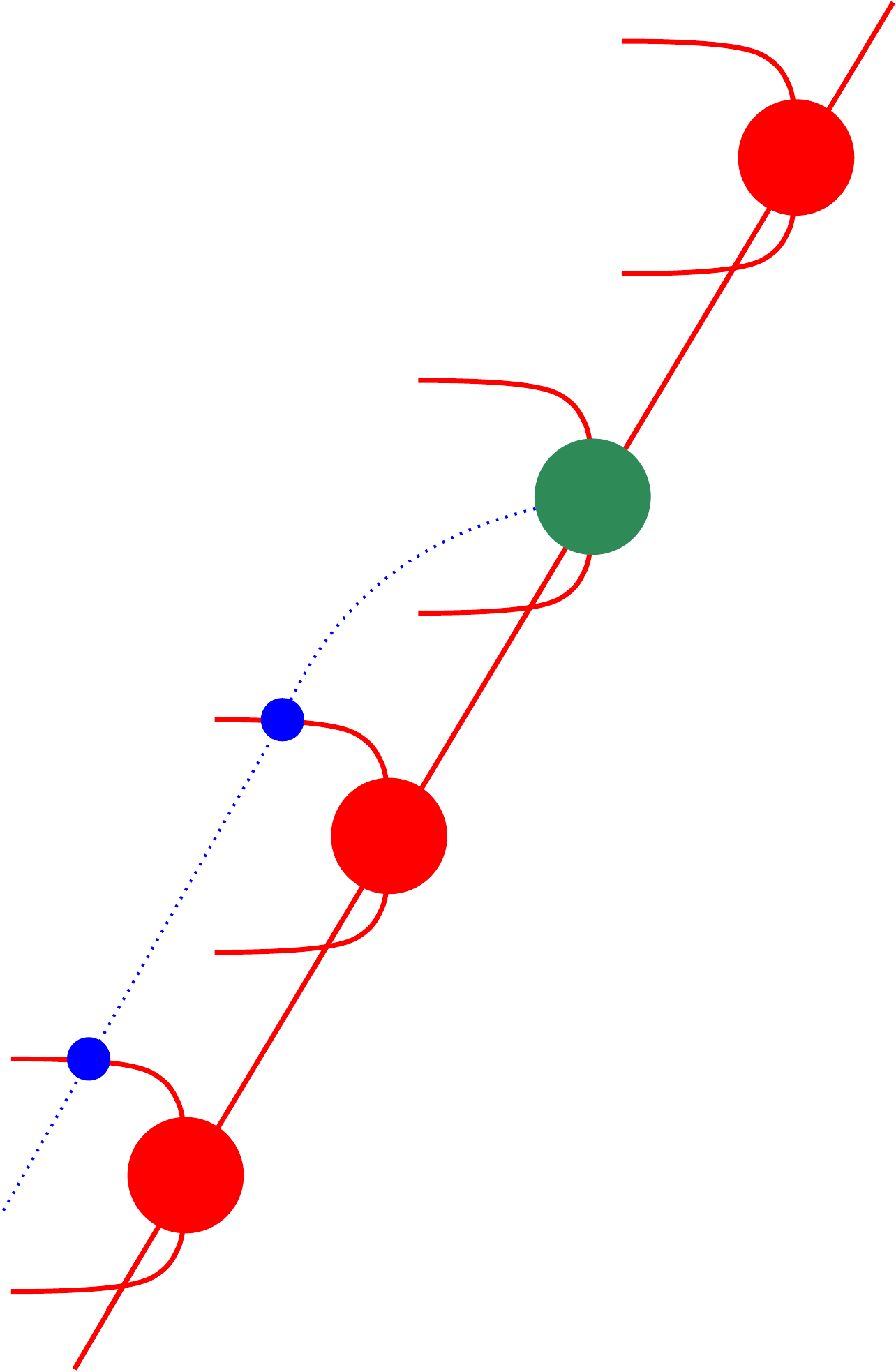}}} \sim
\vcenter{\hbox{
\includegraphics[height=0.2\linewidth]{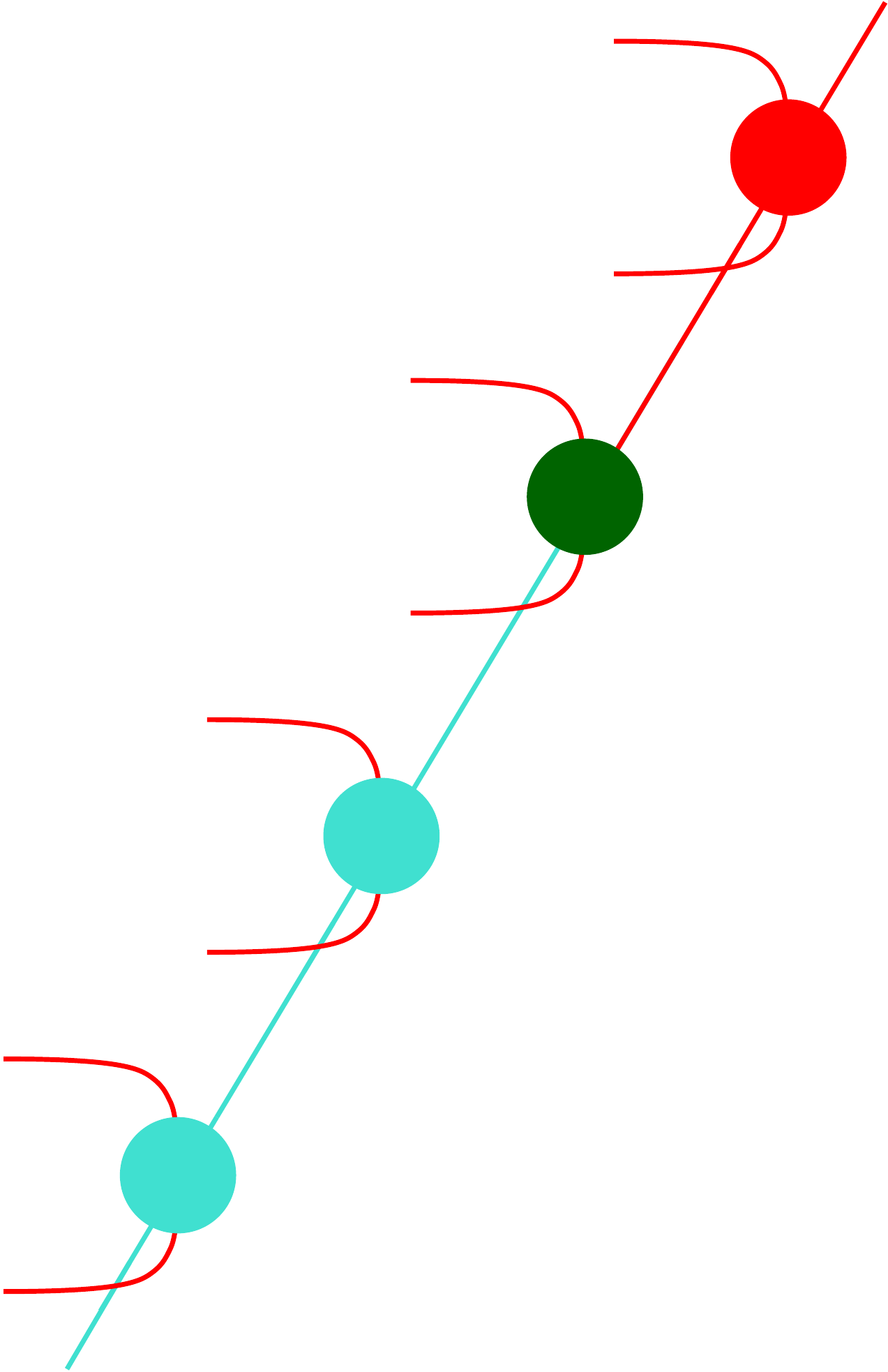}}}
\end{align}
\caption{The ansatz for the topologically trivial excitations of the transfer matrix is obtained by changing one single tensor (green) and making a momentum superposition (left). Topologically non-trivial excitation, corresponding to anyons, are obtained by attaching an MPO string to the perturbed tensor (center). This is equivalent to a kink excitation, where we have a perturbed tensor (green) but different fixed point tensors (red and cyan) on either side of it (right).}\label{fig:excitations}
\end{figure}

We already discussed the fixed points of the transfer matrix. We now look at the other, excited, eigenstates of the transfer matrix. For this, we use the excitation ansatz for 1D systems \cite{haegeman2013elementary}. Given a fixed point, the ansatz for excitations of the transfer matrix is given by locally changing a single tensor and making a momentum superposition. This ansatz can only capture the topologically trivial excitations. There are two alternative but equivalent ways to think about the non-trivial excitation. These excitations correspond to local tensors with an MPO string attached. The relation in Figure \ref{fig:fixedpoint} implies that this is equivalent to the usage of different fixed points to make kink excitations, one on each side of a perturbed tensor. Clearly there is an if and only if relation in this formalism between the existence of excitations with a locally invisible string and the existence of several fixed points related via the application of an MPO. The excitations are graphically denoted as in Figure \ref{fig:excitations}.

\begin{figure}[h!]
\begin{align*}
\braket{\psi[O,k] |\psi[O,k]}=
\vcenter{\hbox{
\includegraphics[height=0.2\linewidth]{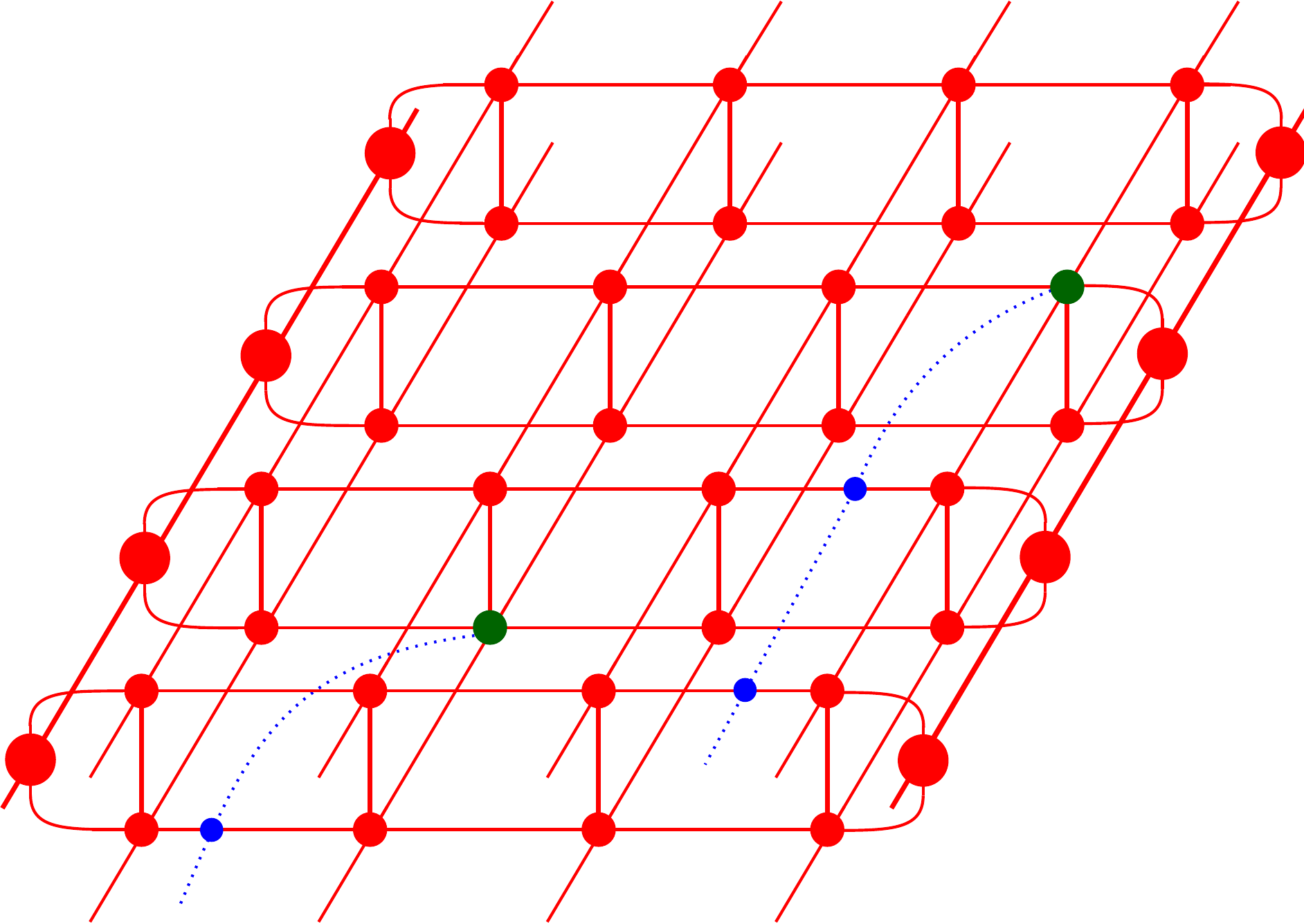}}}=
\vcenter{\hbox{
\includegraphics[height=0.2\linewidth]{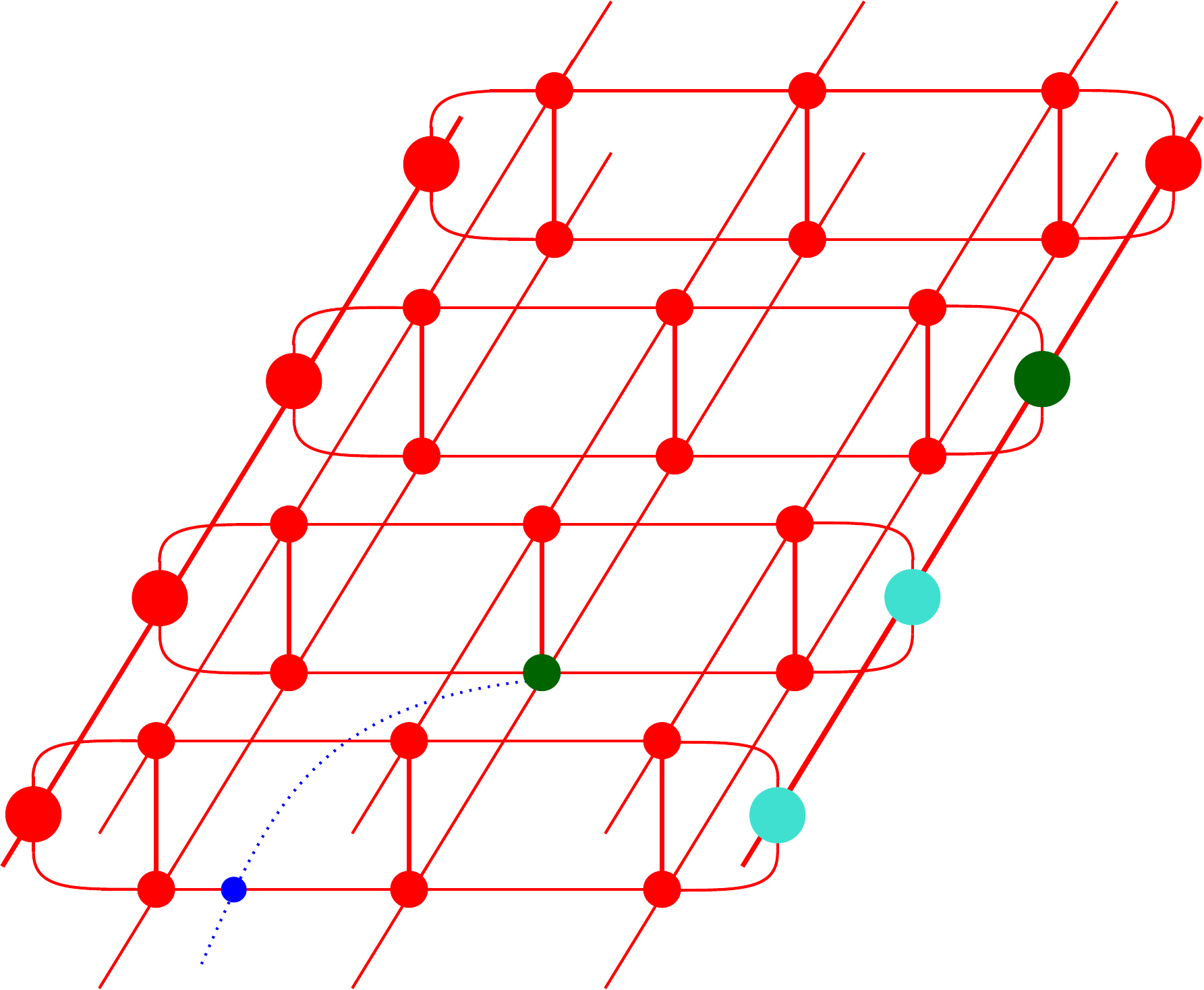}}}=
\vcenter{\hbox{
\includegraphics[height=0.2\linewidth]{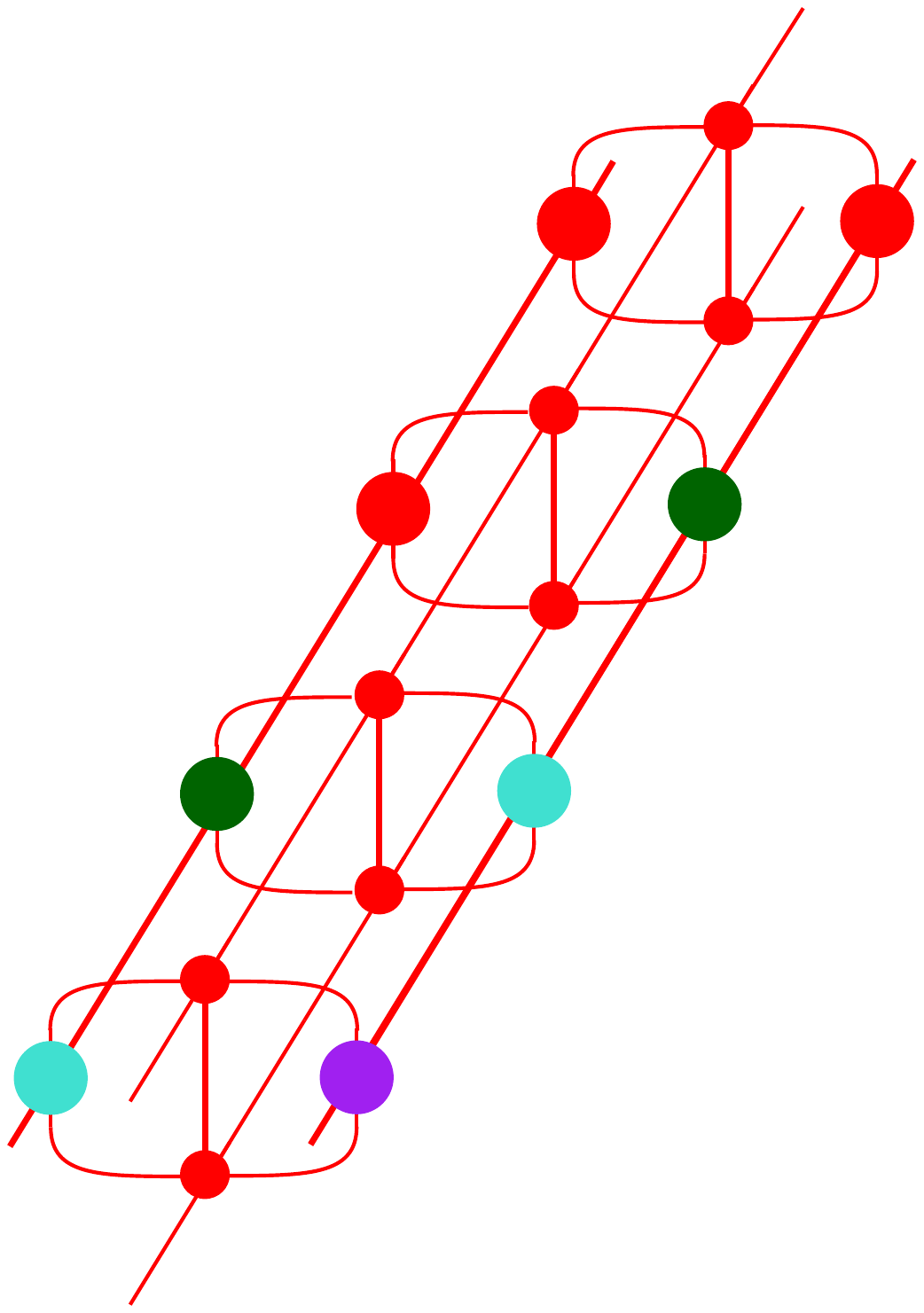}}}
\end{align*}
\caption{Contraction scheme to calculate the norm of an excitation, see the main text for details. The Figure does not show the momentum superpositions for the sake of clarity.}\label{fig:normexcitation2}
\end{figure}

We now return to the original question of calculating the norm of an excitation, see Figure \ref{fig:normexcitationPEPS}. 
The contracting steps are illustrated in Figure \ref{fig:normexcitation2}.  We start contracting this network from the left and right. This gives us the trivial fixed points on the left and right, as they are the unique injective positive fixed points and the transfer matrix is a completely positive map. Physically this corresponds to the fact that at infinity the state is in the topologically trivial ground state. Once we reach the excitation in the ket from the right, we get a kink excitation, with the trivial fixed point on top (red), an excited tensor (green) and another, possibly non-injective fixed point, on the bottom. Contracting the layers between the excitation amounts to applying the transfer matrix, this does not change the fixed points tensors, although it might affect the excited tensor. Next, we reach the position of the excitation in the bra. The application of the MPO gives us a new fixed point (purple). 

Remember that we want to make this quantity as large as possible.  The final outcome contains the local overlap of the red and purple fixed points. If the norm of the state is non zero, the fixed points $\rho_1$ resulting from the application of an MPO in the bra or ket on the trivial fixed point have to contain at least once the same injective fixed point. Written differently, we see that this implies that $O_i \rho_1 O_i^{\dagger} = \rho_1 + \ldots$, or $N_{ii}^1 >0$. This is indeed true at the RG fixed point and should remain so throughout the phase. We see here again that the topological properties of the bulk, the existence of anyons, is through the bulk-boundary correspondence reflected in a symmetry property of the fixed point subspace.  The anyons then correspond to symmetry breaking, domain wall, excitations.

We can thus assume that both fixed points on the bottom of the diagram are the same as this is the only term that contributes to the norm. The computation then depends on what happens to the excited tensor between the location of both fixed points. Because of the momentum superposition this is the application of  $(1-e^{ip_y}T)^{-1}$ to the excited 1D state. Clearly, to maximize this, we want to choose the variational parameters in the green tensor such that the 1D state is an eigenstate of the transfer matrix with eigenvalue as close as possible to $e^{-ip_y}$. As the momenta are variational parameters, we see from the previous discussion that we need to find the eigenvectors of the transfer matrix whose eigenvalues have the largest magnitude.

This calculation again reveals the importance of the symmetries of the fixed point subspace for the topological properties of a system. Because of the pulling through condition, we know that the transfer matrix commutes with the application of an MPO in its ket or bra layer. Hence, the fixed point subspace inherits this symmetry. Generically, a system with this property will have maximal symmetry breaking and (at least) $N^2$ different fixed points, with $N$ the number of MPOs. However, in the PEPS picture we have just argued that $N_{ii}^1 >0$, hence we expect less fixed point, $N$ to be precise, if the system is topologically ordered, in the sense that it has anyonic excitations.

\subsection{Classification of Excitations}
\begin{figure}[h!]
\begin{align*}
\vcenter{\hbox{
\includegraphics[height=0.25\linewidth]{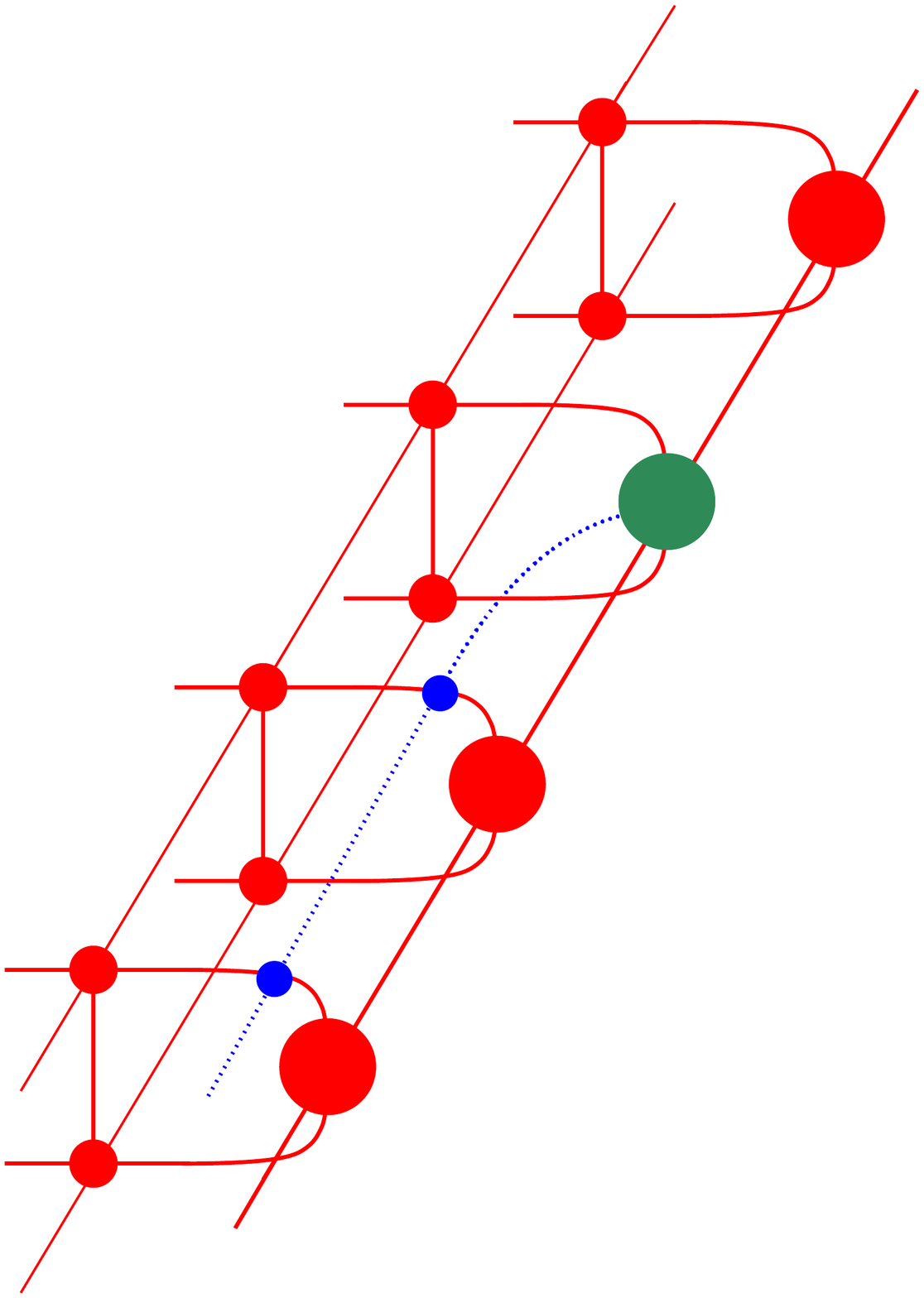}}} = \mu
\vcenter{\hbox{
\includegraphics[height=0.25\linewidth]{Excitation1.pdf}}},\qquad
\vcenter{\hbox{
\includegraphics[height=0.25\linewidth]{EigEquation.pdf}}} =\mu
\vcenter{\hbox{
\includegraphics[height=0.25\linewidth]{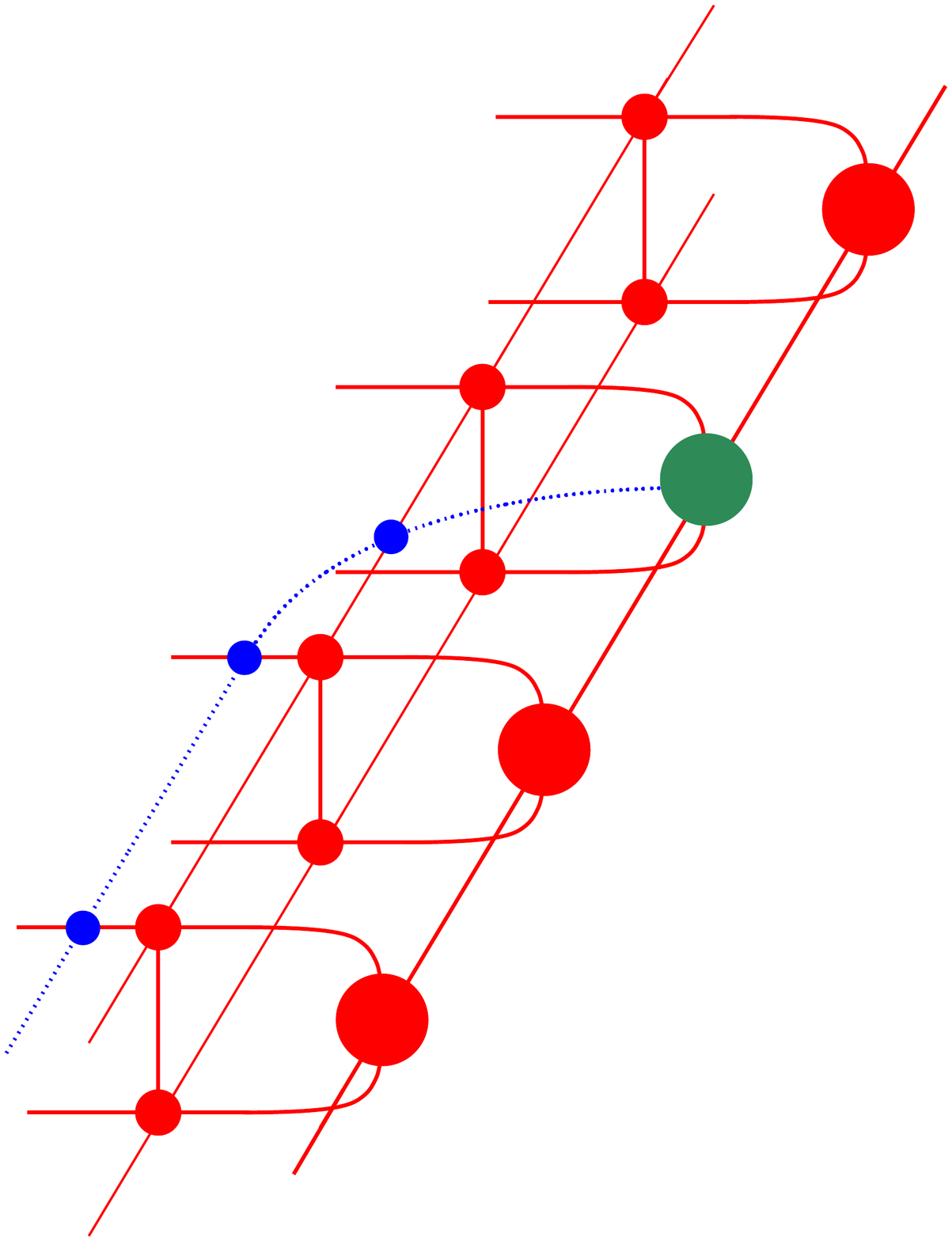}}}
\end{align*} 
\caption{The eigenvector equation for the regular transfer matrix (left). An equivalent equation, using the pulling through property (right).}
\label{fig:eigenvalueT}
\end{figure}

\begin{figure}[h!]
\begin{align*}
\vcenter{\hbox{
\includegraphics[height=0.25\linewidth]{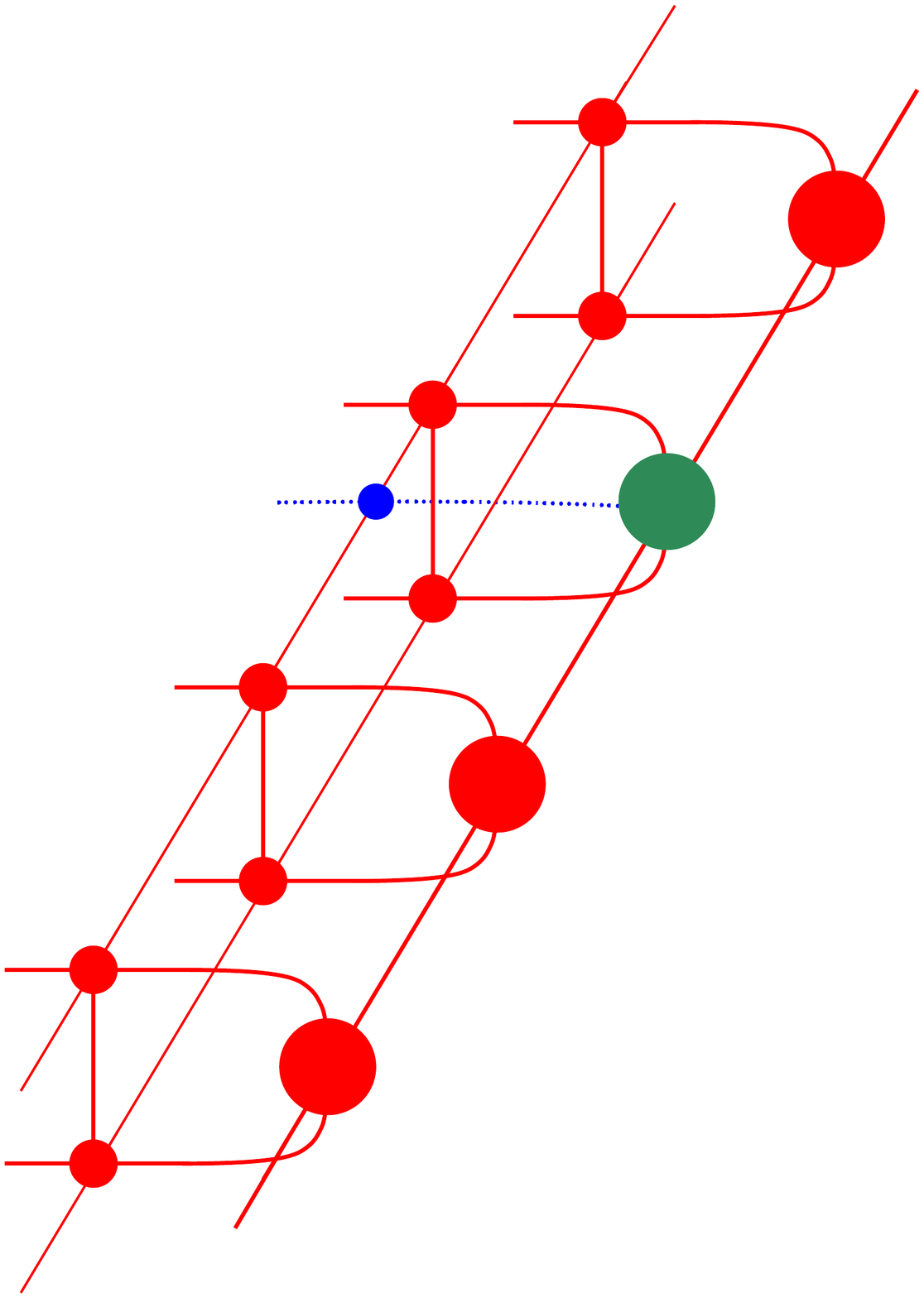}}} = \mu
\vcenter{\hbox{
\includegraphics[height=0.25\linewidth]{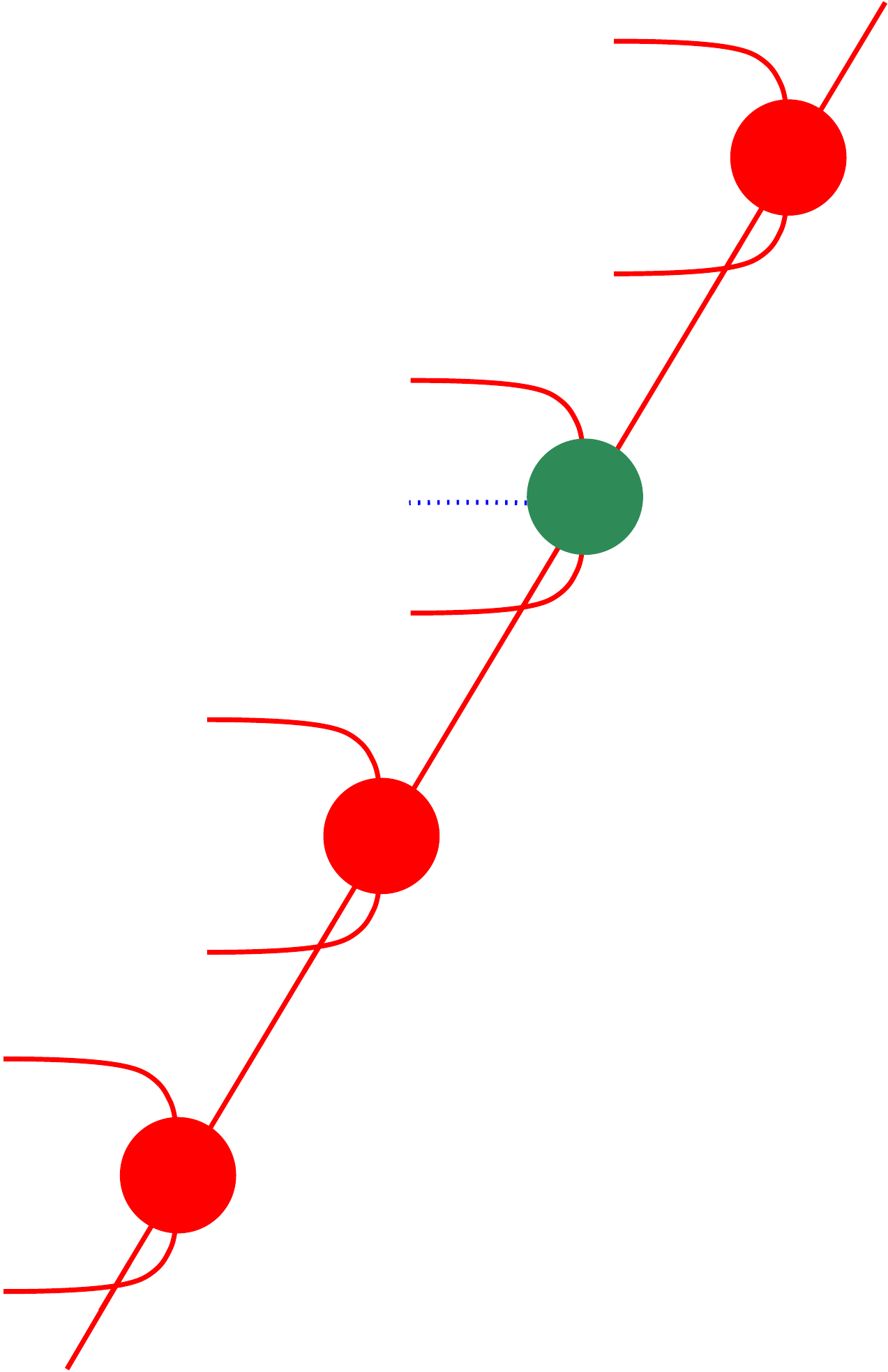}}}, \qquad \qquad
\vcenter{\hbox{
\includegraphics[height=0.25\linewidth]{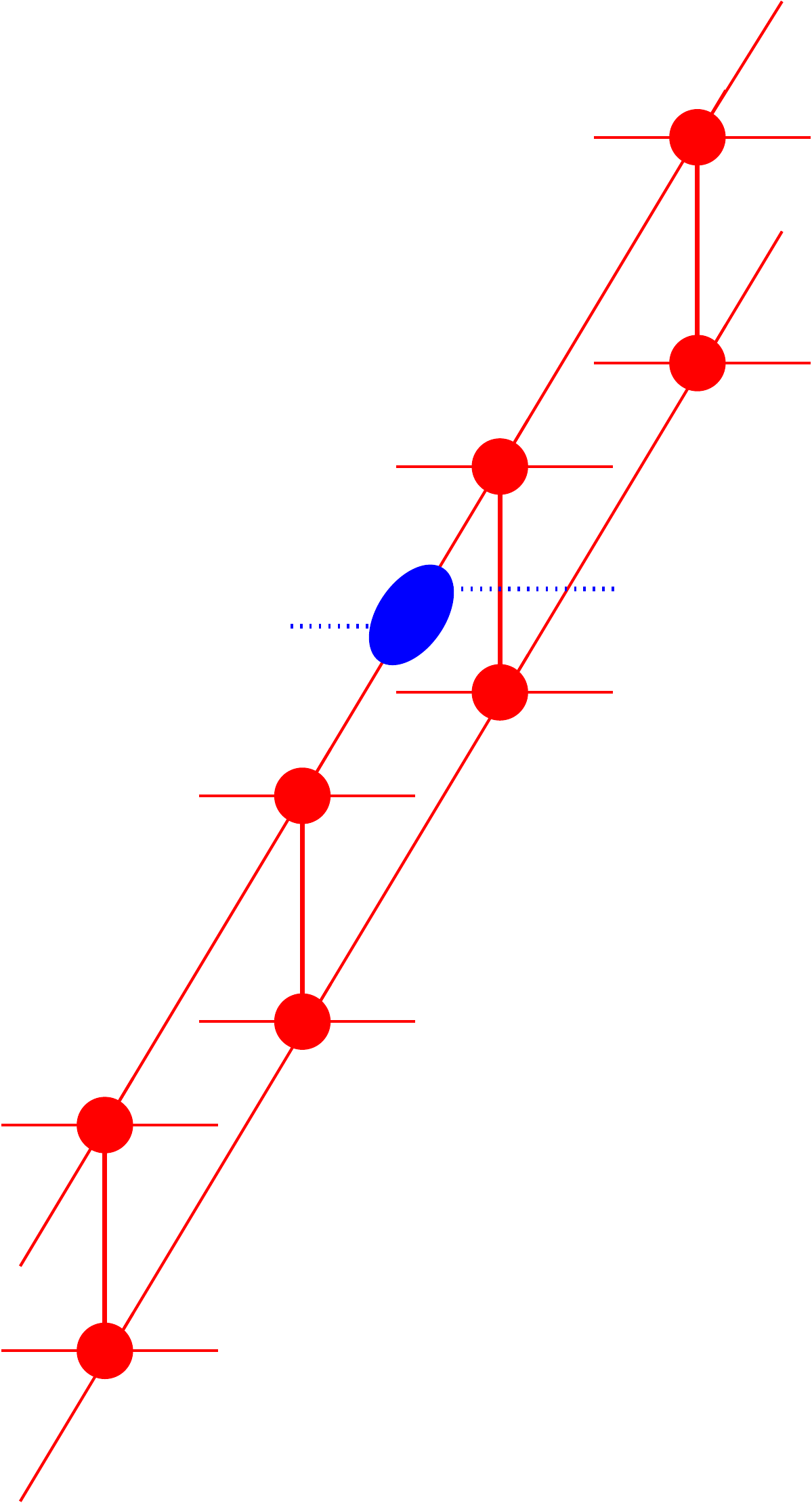}}}
\end{align*} 
\caption{The eigenvalue equation for the mixed transfer matrix (left). This is the regular transfer matrix with a extra MPO through its ket (or bra) layer. The excitation ansatz consists now of  local perturbations on the trivial fixed points, but the perturbed tensors (green) have en extra MPO index. We can naturally use the idempotents to define operators in the mixed formalism (right).}
\label{fig:eigmixedT}
\end{figure}

We now discuss how we can classify the excitations of the transfer matrix with the appropriate anyon labels of the original two dimensional topological theory. As discussed an excitation of the transfer matrix such as in Figure \ref{fig:normexcitation2} actually encodes the existence of an excitation with a string in the right half of the full 2D model. We can measure the anyon type of a part of the lattice using the central idempotents. To apply these to the 1D excitations we need some extra manipulations we now explain. Because of the pulling through condition we can rewrite the eigenvalue equation as in Figure \ref{fig:eigenvalueT}. It is then clear that we can equivalently look for eigenvectors of the so called mixed transfer matrix that contains an extra MPO in the ket, see Figure \ref{fig:eigmixedT}.

\begin{figure}[h!]
\begin{align*}
\vcenter{\hbox{
\includegraphics[height=0.25\linewidth]{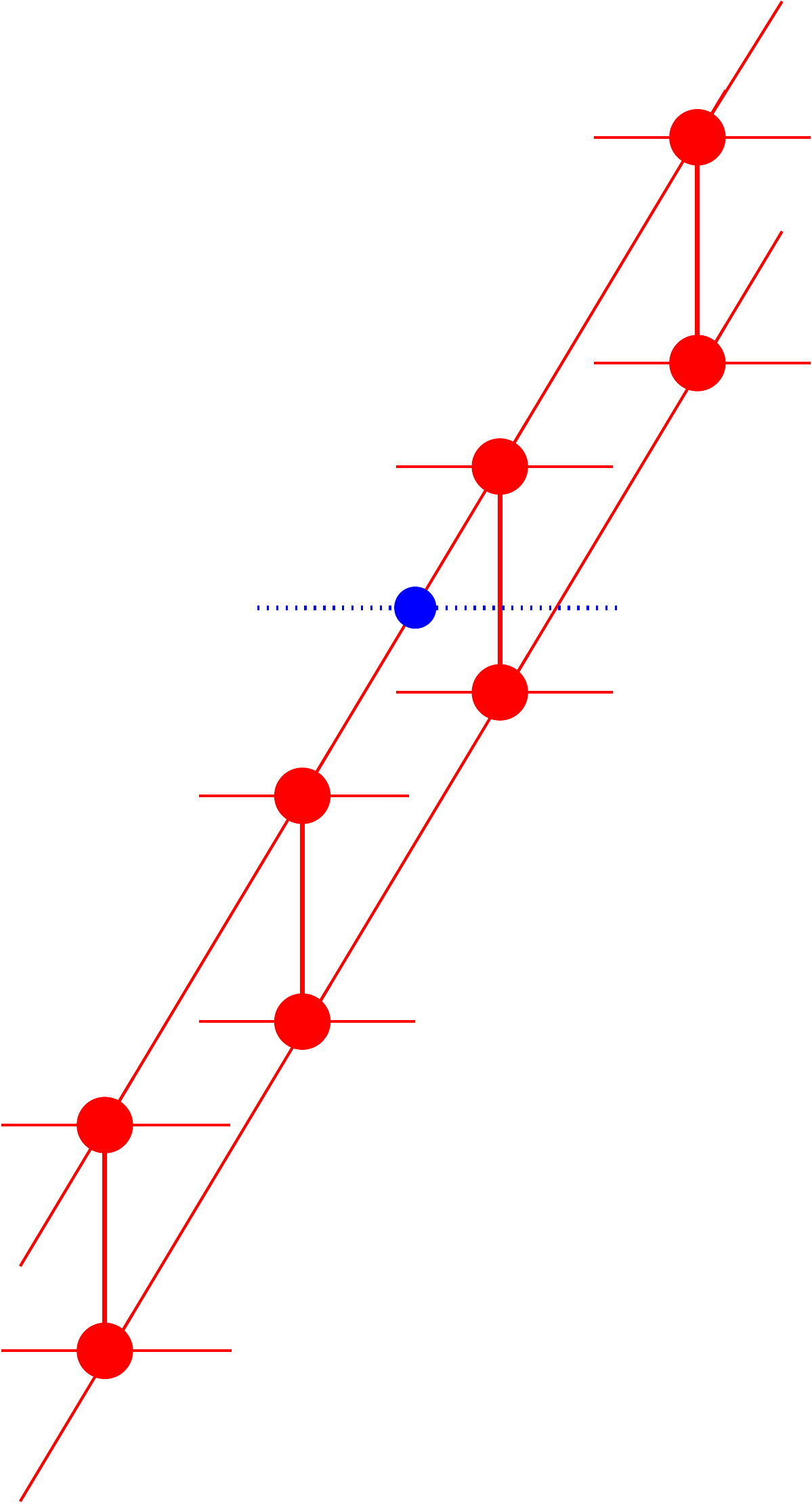}}}=
\sum_i\vcenter{\hbox{
\includegraphics[height=0.25\linewidth]{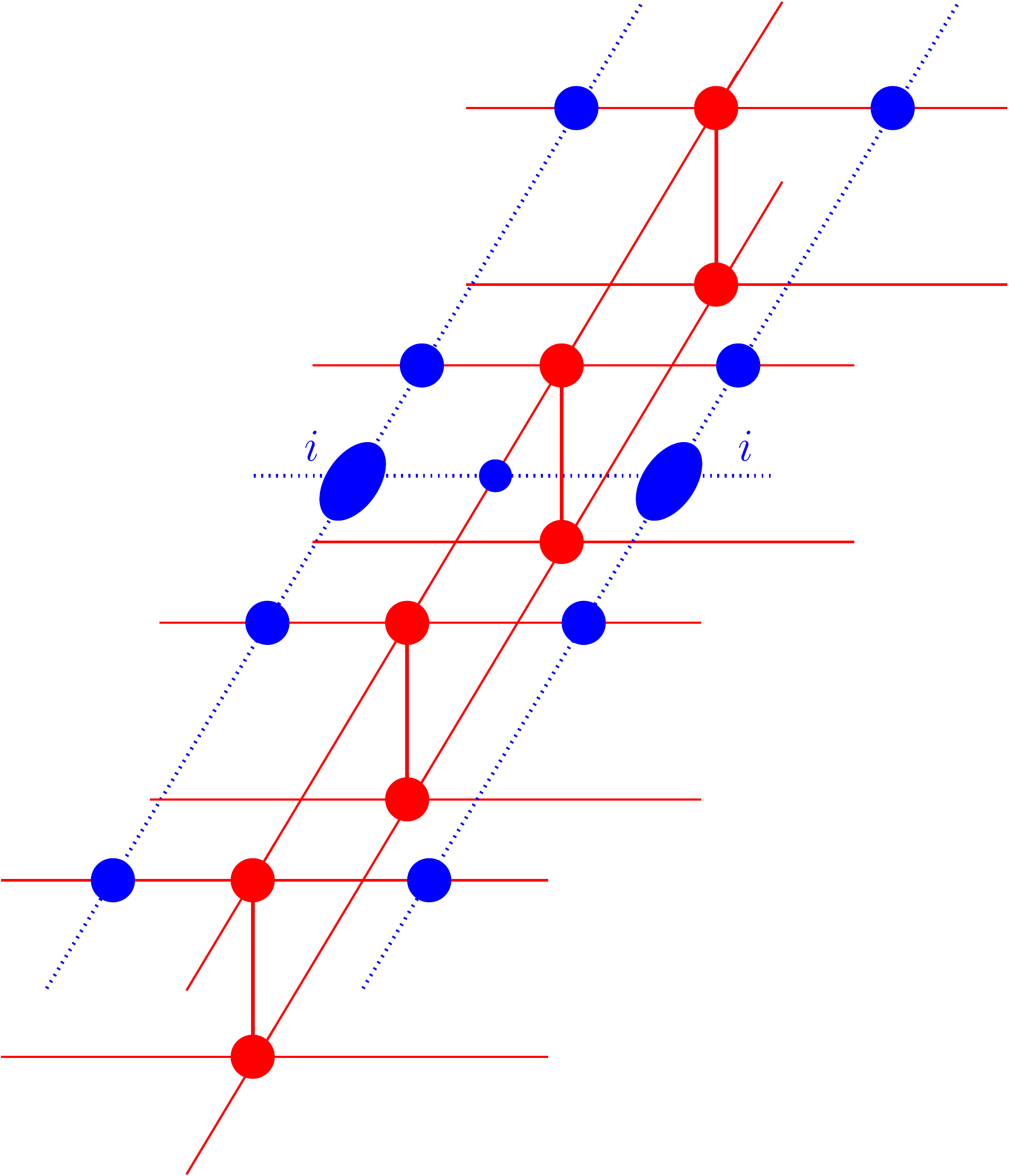}}}.
\end{align*} 
\caption{The mixed transfer matrix has a block decomposition induced by all the idempotents.}
\label{fig:mixedTdeco}
\end{figure}

The reason we did this is to have access to the blue MPO label, which we need to apply the central idempotents. The mixed transfer matrix also reveals the extra topological structure more clearly. Indeed, as was shown in \cite{bultinck2015anyons} all idempotents commute with the mixed transfer matrix, giving rise to a block decomposition into the different topological sectors. It is this decomposition that makes the classification of excitations possible. This property is illustrated in Figure \ref{fig:mixedTdeco}. A similar property holds when we apply the idempotents in the bra instead of the ket layer.

\begin{figure}[h!]
\begin{align}
\braket{\psi[O,k]|\mathcal{P}_i|\psi[O',k]}=
\vcenter{\hbox{
\includegraphics[height=0.2\linewidth]{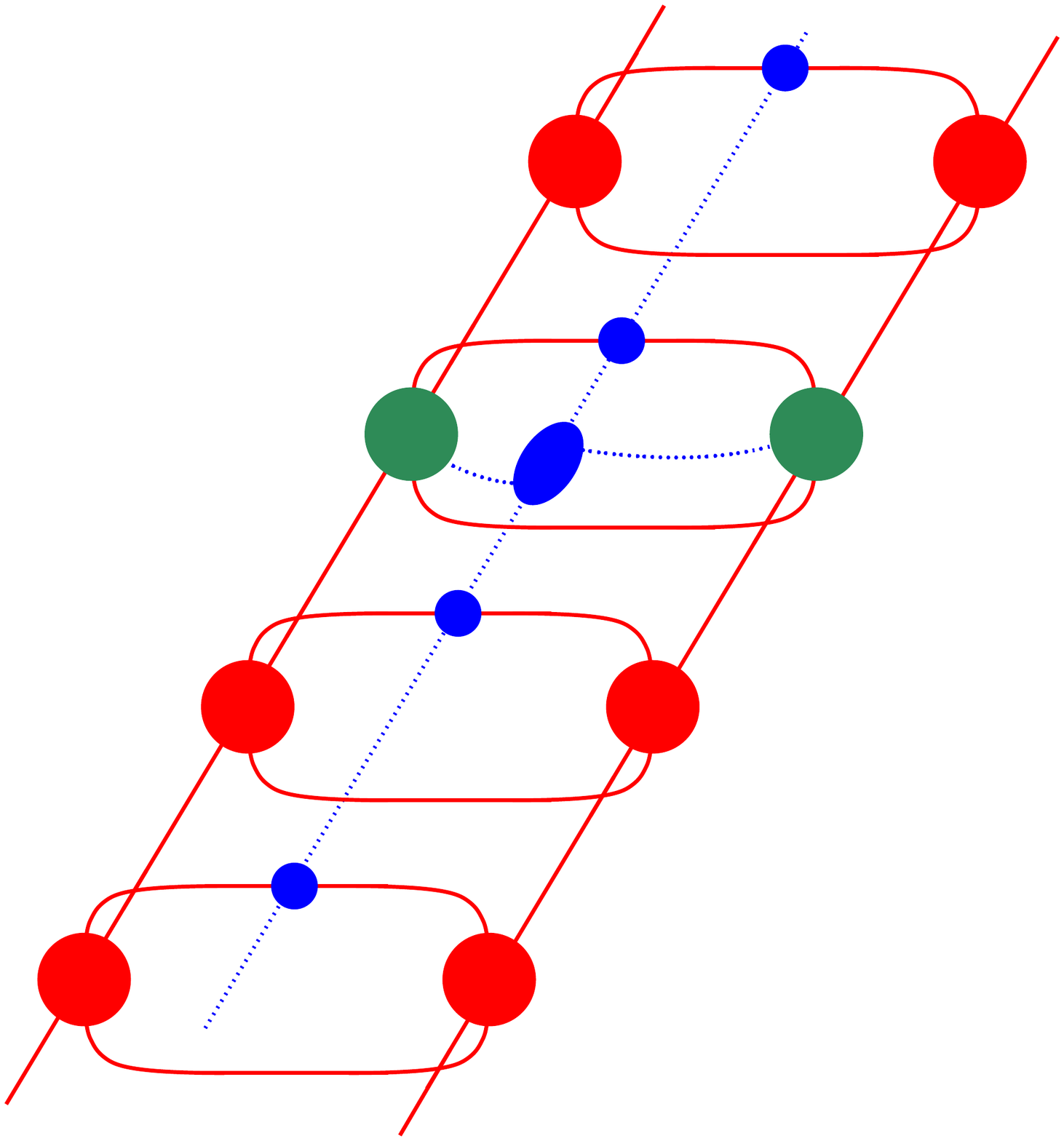}}}.
\end{align}
\caption{The network that gives the matrix element of a central idempotent with respect to some excitations at a given energy and momentum. As this is a 1D network it can be contracted efficiently.}
\label{fig:expvalueIdem}
\end{figure}

For the classification it suffices to consider only one of the cases. We only apply non trivial idempotents in the ket, in the bra we will not place an extra idempotent, or equivalently, always place the idempotent corresponding to the topologically trivial excitations in the bra. We can now proceed as follows. We gather all excitations, both trivial and kink, of the transfer matrix that correspond to the same eigenvalue and momentum. We list all the possible ways we can  explicitly obtain an MPO string from such an excitation as in Figure \ref{fig:excitations}. If the theory is non-Abelian, there can be several possibilities. We can now calculate the matrix elements of all the central idempotents projected onto this eigenspace using standard MPS methods and diagonalize the result as in Figure \ref{fig:expvalueIdem}. All the idempotents are still projectors in this subspace, due to the block structure of the transfer matrix. The central idempotents whose restrictions have non zero rank give the anyon types of the excitations in the considered energy and momentum spaces.

\end{document}